\documentclass[final]{IEEEtran}
\IEEEoverridecommandlockouts

\usepackage{times}
\usepackage[final]{graphicx}
\usepackage{amsmath,amsfonts}
\usepackage{amsthm}
\usepackage{amssymb,amsbsy}

\usepackage{cite}
\usepackage{multirow}

\usepackage{color}

\usepackage{tabularx, booktabs}
\usepackage{bigstrut}
\usepackage{booktabs}
\usepackage{rotating}

\newcommand{\ignore}[1]{}

\newcommand{\hsp}{\hspace{0.1in} }
\newcommand{\hspp}{\hspace{0.05in} }
\newcommand{\hsppp}{\hspace{0.02in} }

\newsavebox{\savepar}

\begin{document}

\title{Millimeter Wave Channel Measurements and Implications for PHY Layer Design}
\author{Vasanthan Raghavan, Andrzej Partyka, Lida Akhoondzadehasl,
Ali Tassoudji, Ozge Koymen, and John Sanelli \\
Qualcomm Corporate R\&D, USA \\
E-mail: \{vraghava, apartyka, lidaa, alit, okoymen, jsanelli\}@qti.qualcomm.com
\thanks{An initial version of this paper was published at the IEEE Global Telecommunications
Conference, San Diego, CA, Dec.\ 2015~\cite{ozge}.}}

\maketitle

\begin{abstract}
\noindent
There has been an increasing interest in the millimeter wave (mmW) frequency
regime in the design of next-generation wireless systems. The focus of
this work is on understanding mmW channel properties that have an important
bearing on the feasibility of mmW systems in practice and have a significant impact
on physical (PHY) layer design. In this direction, simultaneous channel sounding
measurements at $2.9$, $29$ and $61$ GHz are performed at a number of
transmit-receive location pairs in indoor office, shopping mall and outdoor
environments. Based on these measurements, this paper first studies large-scale
properties such as path loss and delay spread 
across different carrier frequencies in these scenarios. Towards the goal of
understanding the feasibility of outdoor-to-indoor coverage, material measurements
corresponding to mmW reflection and penetration are studied and significant notches in signal
reception spread over a few GHz are reported. 
Finally, implications of these measurements on system design
are discussed and multiple solutions are proposed to overcome these impairments.
\end{abstract}




\section{Introduction}
\label{sec1}
Recent attention on millimeter wave (mmW) systems for meeting the high data-rate
demands in next-generation devices has resulted in a burgeoning interest in the
focus on these systems~\cite{khan,qualcomm,rappaport,boccardi1}. Given the
unfavorable wireless propagation in mmW settings~\cite{5G_whitepaper,3gpp_CM_rel14},
it is now understood that physical (PHY) layer system design has to consider
innovative approaches to address these impairments. In particular, beamforming with a
large number of antennas~\cite{rusek,hur,roh,sun,brady,oelayach,raghavan_jstsp,rangan,ghosh}
is a key ingredient in meeting the link margin of mmW systems. Towards the
goal of realizing beamforming-based mmW systems, there has been a strong interest
in understanding radio frequency (RF) design challenges for large bandwidth
systems~\cite{rebeiz,larson,harish}, as well as measurements and channel modeling at
different carrier frequencies of interest. In particular, there have been multiple
studies in the modeling of $60$ GHz indoor channels~\cite{hao_xu,zwick,smulders}, as
well as channels at
other carrier frequencies such as $15$ GHz~\cite{okvist},
$28$ GHz~\cite{samimi_conf,azar,zhao}, $38$-$39$ GHz~\cite{rappaport_icc},
$73$ GHz~\cite{maccartney}, etc.

The scope of this work is on further understanding some of the representative
large-scale mmW channel characteristics, additional impairments encountered in
practical implementations of these systems, and the implications these observations
have on PHY layer design.

Towards this goal, we start with channel propagation
comparisons based on measurements at the {\em \underline{same}} transmit-receive
location pairs at $2.9$, $29$ and $61$ GHz in the indoor office, shopping mall and
outdoor environments. In addition to the
vast set of measurements in different settings reported here, this work makes an
important and novel contribution given the fact that most papers in the mmW
literature consider channel measurements at individual mmW frequencies and not at
the same location pairs. Such a study allows us to legitimately compare propagation
at different frequencies.
Our studies show that losses at mmW
frequencies are typically higher than with sub-$6$ GHz systems in both indoor and
outdoor settings. However, these losses are not significantly worse at the mmW
regime relative to sub-$6$ GHz settings. While the observed delay spreads are
typically small in indoor settings, outdoor settings can lead to significant
increases in delay spreads corresponding to strong reflections from distant objects.
Beamformed delay spreads are expected to be smaller than omni delay spreads in most
scenarios.

Building on these studies, we then branch off into impairments arising from
specific practical implementations. Given that indoor WiFi replacement is an
important use-case of mmW systems and actively considered by the 5GTF
(${\tt www.5gtf.org}$), we then study the impact of reflection response and
penetration through
walls of residential buildings. Our studies show that significant loss
of coverage can be observed at mmW frequencies corresponding to deep frequency
notches which needs to be overcome with adequate spatial and frequency diversity.
We also report that at mmW frequencies, the penetration depth of electromagnetic
(EM) energy into the human body is small and a significant fraction of the energy
is reflected. This observation is of particular importance in stadium deployments.

\ignore{
Given the diminishing $3$-dB beamwidth of the directional beamforming vectors
with antenna dimensions~\cite{oelayach,raghavan_jstsp}, mmW systems are
susceptible to blockages of signals much more than sub-$6$ GHz systems are.
In particular, mmW systems are susceptible to self-blockage, which is shadowing
from the user itself in the form of hand blocking and blockage from other body
parts. This can cause a complete blockage of the user equipment (UE) antennas
depending on the antenna position relative to the hand. In addition, there are
other blockages from the environment around the UE in the form of humans, vehicles,
foliage and other objects. With this background, we then focus on blockage issues
in this paper~\cite{ozge,zhao_ericsson,5G_whitepaper,3gpp_CM_rel14}.

To model self-blockage, EM simulations of antennas in the proximity of the hand
are performed and angular regions corresponding to signal blockage with
different user grips are identified. To model other blockages, simulation
studies are conducted to capture the impact of objects at the UE in the form of
angular regions blocked and losses incurred with a knife-edge diffraction model.
These studies lead to a statistical blockage model that has many attractive
properties: i) parsimonious (captured by a small number of model parameters),
ii) efficacious (captures the real impact of blockages), and iii) computationally
efficient (easily useable in a system simulator framework) in studying the
performance of mmW systems.
}

This paper is organized as follows. Section~\ref{sec2} provides a brief overview
of the channel sounder, measurement methodology and measurement scenarios.
Section~\ref{sec3} studies large-scale channel parameters such as path loss and
delay spreads 
in indoor and outdoor environments. Section~\ref{sec4} reports reflection response and
penetration loss characterization with materials
commonly found in residential and indoor environments. 
Section~\ref{sec6} illustrates the impact on PHY layer of all
the observations from Secs.~\ref{sec3}-\ref{sec4} and Section~\ref{sec7} concludes
the paper.

\section{Measurement Setup and Methodology}
\label{sec2}

\subsection{Channel Sounder}
\label{sec2a}
We begin with a brief description of the channel sounder and measurement methodology.
Measurements were performed 
with a battery-powered and freely-mobile channel sounder that allows automatic
omni-directional scans at $2.9$, $29$ and $61$ GHz, and elevation and azimuthal
scans at $29$ and $61$ GHz. Parallel data-sets for these frequencies were obtained
at identical transmit and receive locations using omni-directional antennas without
swapping cables.
In addition, directional horn antennas with $10$ and $20$ dBi gains were used
to obtain measurements at $29$ and $61$ GHz. Directional
scans consisted of azimuthal ($360^{\sf o}$ view) and spherical scans ($360^{\sf o}$
azimuth view and $-30^{\sf o}$ to $90^{\sf o}$ view in elevation). The resultant scans
included $39$ slices with a $10$ dBi gain antenna and $331$ slices with a $20$ dBi
gain antenna.
The resolution of the channel sounder is approximately $5$ ns.
An Agilent E8267D signal generator is used to generate a pseudo-noise (PN)
sequence at a chip rate of $100$ Mc/s, which is then used to sound the channel. At
the receiver, an Agilent N9030A signal analyzer is used for acquisition and the PN
chip sequence is despread using a sampler at $200$ MHz and with $16$ bit resolution.

The omni-directional antennas allow us to measure path loss variations up to $180$ dB.
Thus, in understanding macroscopic channel properties (such as path loss and delay
spread), we do not need to obtain the omni-directional channel response by stitching
together directional/horn responses. However, our measurements are constructed out of
a variable-time integration of omni-directional responses to ensure that there is sufficient
signal strength to allow signal discrimination. This variable time could be anywhere from
less than $1$ ms to a few tens of ms, depending on the link margin/distance between
transmitter and receiver. Up to $80$ dB processing gain is realized in the channel
acquisition process.

\begin{figure*}[htb!]
\begin{center}
\includegraphics[height=3.8in,width=6.15in] {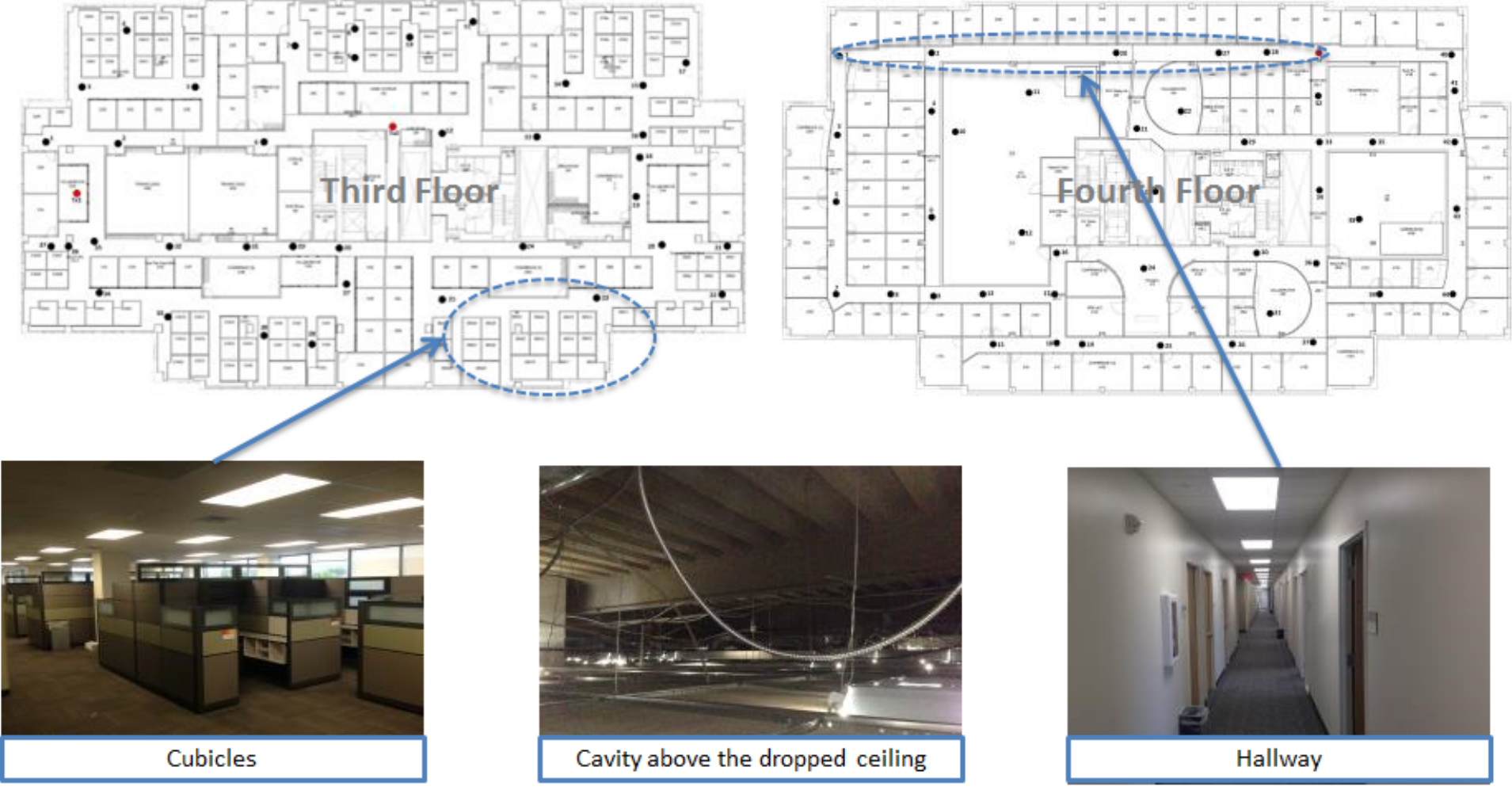}
\caption{
Indoor layout across the third and fourth floor of the Qualcomm building.
Transmitter and receiver locations are marked in red and black circles, respectively.
}
\label{fig_layoutsa}
\end{center}
\vspace{-5mm}
\end{figure*}

\begin{figure*}[htb!]
\begin{center}
\begin{tabular}{cc}
\includegraphics[height=2.1in,width=3.15in] {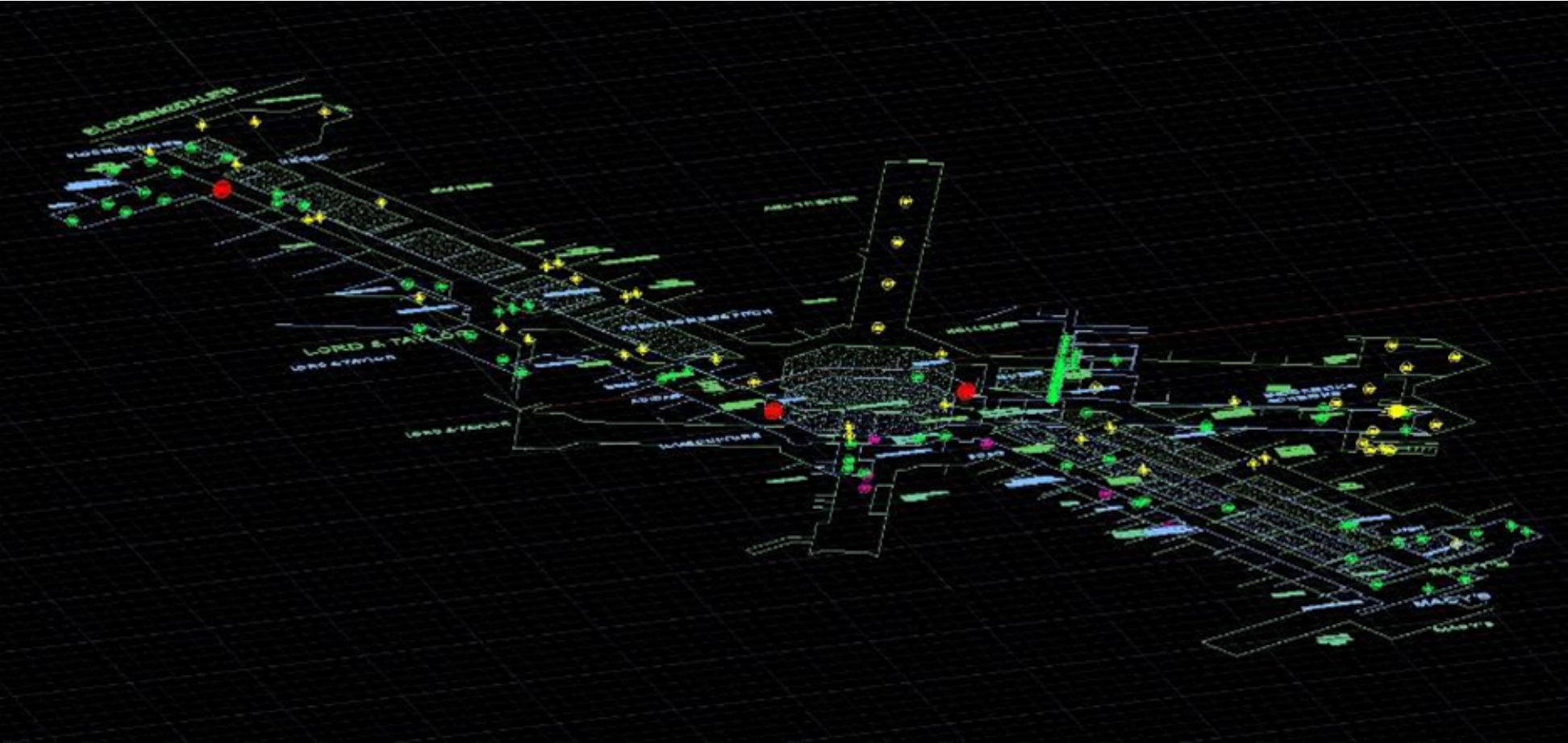}
&
\includegraphics[height=2.1in,width=2.65in] {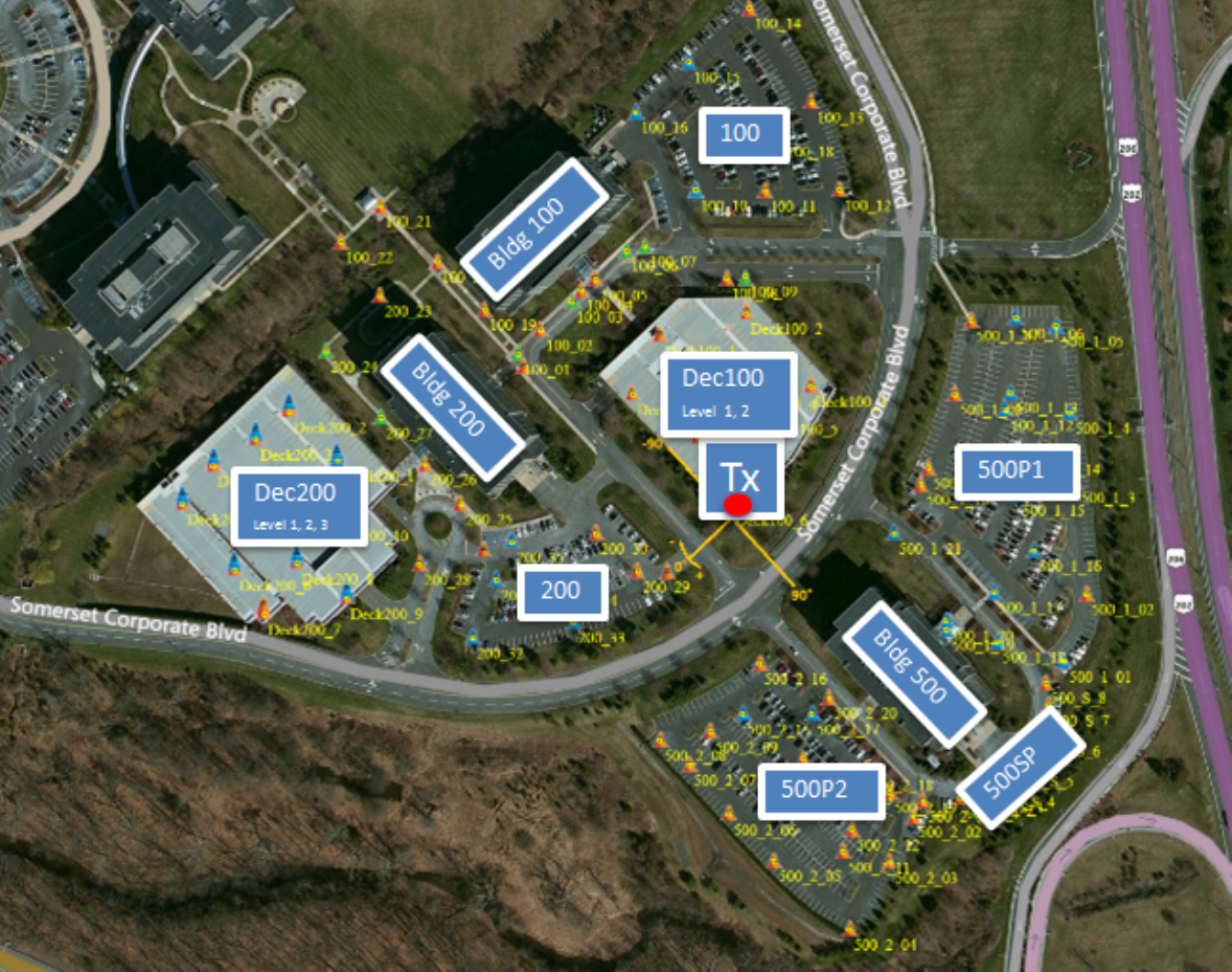}
\\
(a) & (b)
\end{tabular}
\caption{(a) Shopping mall layout with transmitter and receiver locations
in red and green/yellow circles. (b) Measurement layout outside the Qualcomm
building in Bridgewater, NJ.}
\label{fig_layouts}
\end{center}
\vspace{-5mm}
\end{figure*}

\subsection{Measurement Locations}
\label{sec2b}
Indoor office measurements were made across two floors of the Qualcomm building in
Bridgewater, NJ, USA, with dimensions of $75 \times 40 \times 2.68$ m$^3$. The
building construction is representative of a modern office building in the USA.
See Fig.~\ref{fig_layoutsa} for more details on the office layout across the two
floors and measurement locations (transmitter in red and receiver in black).

The two floors represent two types of typical office environments. The third floor is
mostly comprised of cubicles along the edge of the floor plan with walled offices and
conference rooms towards the center. The fourth floor is comprised of walled offices
(larger than the third floor), conference rooms and laboratories. The partition walls
are constructed with metallic studs spaced at $0.46$ m ($1.5$ ft) intervals. The
ceiling is a dropped ceiling $2.7$ m ($\sim {\hspace{-0.04in}} 9$ ft) above the floor
with an additional $0.91$ m ($\sim {\hspace{-0.04in}} 3$ ft) cavity below the concrete
ceiling. While the cavity is not a visible aspect of the office, the abundance of
metal objects such as concrete ceiling with a corrugated metal substrate and metal
duct-work pipes in a fairly open space plays an important role in propagation measurements.
On the third floor, the measurements were made between two transmitter
locations\footnote{Only $2.9$ and $29$ GHz measurements were done at the second transmitter
location due to logistical constraints.} and the same set of multiple receiver locations.
The first transmitter location is centrally located, while the second one is positioned
at the left-hand edge of the floor plan. On the fourth floor, the measurements were made
between a single transmitter location and multiple receiver locations. Given the high
density of partition walls in the office building, a large majority of the
measurements were non-line-of-sight (NLOS) in nature.

Indoor shopping mall measurements were made at the Bridgewater Commons Mall, Bridgewater,
NJ, USA, which is a large three level indoor shopping mall with an open interior design.
The mall layout as well as the measurement locations are seen in Fig.~\ref{fig_layouts}(a).
The building length is $\sim{\hspace{-0.04in}}390$ m with the longest testing range of
$\sim{\hspace{-0.04in}}275$ m. Measurements were obtained at three transmitter and
multiple receiver locations (on all the three levels of the mall). The transmitter
locations were: i) centrally located on the second floor, ii) located on an edge of
the second floor, and iii) centrally located on the third floor near an open-area
food court. Multi-floor propagation was also studied. The specific design of the mall
leads to the observation of a number of both line-of-sight (LOS) and NLOS links.

The first set of outdoor measurements were obtained in downtown New Brunswick, NJ
corresponding to an Urban Micro (UMi)-type environment. Measurements were obtained
from one transmitter location (antenna height of $6.5$ m) and multiple receiver
locations (all at a height of $1.5$ m).
The second set of outdoor measurements were obtained outside the Qualcomm building
and around Somerset Corporate Boulevard in Bridgewater, NJ (see layout in
Fig.~\ref{fig_layouts}(b)). The measurement site is
mostly a tree-lined open square-type setting with some street canyon-type environment.
Specific points of interest include parking lots and structures with bordering buildings,
vegetation\footnote{We point the readers to references such as~\cite{wang2,jones_report,ITU_vegetation}
that estimate the attenuation through vegetation/foliage at mmW frequencies.}
which is a mix of pine 
and spruce 
trees, and a large shopping
mall in close vicinity (Bridgewater Commons Mall). Highways (US Rt.\ $22$ and
I-$287$) are in close proximity with occasional reflections from vehicles in a number
of measurement scenarios. The measurements were made from a single transmitter location
and multiple receiver locations from within $10$ points-of-interest, with
transmit-receive link distances ranging from $35$-$256$ m. Of particular interest
are links corresponding to open areas and parking structures\footnote{In the parking
structures, $61$ GHz measurements were not obtained due to logistical constraints.}.


Some indoor office measurements were obtained over regular day-time office hours and some
were obtained over non-office hours. Due to logistical constraints, shopping mall
measurements were conducted over the night-time with minimal footfall and in common
areas with no inside store access. Tests in downtown New Brunswick were performed during
the day-time with a heavy pedestrian and vehicular traffic in the measurement areas.
Measurements outside the Qualcomm building were done during the day-time with sporadic
(and mostly)
vehicular traffic in the neighborhood. Our measurements do not indicate a significant
impact of humans or vehicles on macroscopic metrics such as path loss and delay spread.
This is because our measurements were repeated $5$ times at the same location to
average over dynamic influences. Each receiver location was measured at five separate
locations offset from each other (center of a circle with a radius\footnote{The
$18$ cm radius corresponds to a fixed railing on whose circumference the horn antenna
is placed for spatial sampling.} of $18$ cm along with $4$ points on the circumference)
to get diversity in measurements and to avoid possible
nulls in the channel response. The correlation over the temporal measurements was high
(over $90 \%$) indicating that the scale at which humans move did not have a substantial
effect on the macroscopic channel properties.

\ignore{
The transmit and receive antennas were placed 10 inches
(24.4 cm) and 29 inches (73.4 cm) below the dropped ceiling,
respectively, to avoid any interaction with measurement
equipment. For each captured channel response, a maximum of
15 samples were classified as legitimate paths. The minimum
power requirement for such classification was set to be min(12
dB SNR, 20 dB below the largest path) where the SNR is
referenced post processing gain.
}

\section{Large-Scale Channel Properties}
\label{sec3}

\begin{table*}[htb!]
\caption{Path Loss Coefficients for the Different Indoor Settings.
$f_{\sf c}$ is in GHz, $\sigma_{\sf CI}, \sigma_{\sf ABG}, \beta'$ are in ${\rm dB}$.}
\label{table1_indoor}
\begin{center}
\begin{tabular}{|l
|c|| c|c|c|c|c|c|| c|c|c|c|c|c|}
\hline
\multicolumn{14}{|c|}{${\bf  {\hspace{0.5in}}
Indoor \hspp Office}$}
\\
\hline
&
& \multicolumn{6}{c||} { ${\sf LOS}$ } & \multicolumn{6}{c|} { ${\sf NLOS}$}
\\ \cline{1-14} 
&
$f_{\sf c}$ 
& ${\sf No. meas.}$
& ${\sf PLE}$
& $\sigma_{\sf CI}$ 
& $\alpha$
& $\beta'$
& $\sigma_{\sf ABG}$ 
& ${\sf No. meas.}$
& ${\sf PLE}$
& $\sigma_{\sf CI}$ 
& $\alpha$
& $\beta'$
& $\sigma_{\sf ABG}$ 
\\ \hline 
\multirow{3}[6]{0.3cm}{\rotatebox[origin=c]{90}{{\hspace{0in}} {\sf All}}} &
$2.9$ & $17$ & $1.62$ & $5.49$ & $2.11$ & $35.47$ & $5.28$
& $105$ & $3.08$ & $6.60$ & $4.36$ & $23.08$ & $5.81$
\\ \cline{2-14}
&
$29$ & $17$ & $1.46$ & $4.25$ & $1.48$ & $61.36$ & $4.25$
& $106$ & $3.46$ & $8.31$ & $4.96$ & $39.79$ & $7.45$
\\ \cline{2-14}
&
$61$ & $19$ & $1.59$ & $4.81$ & $1.03$ & $75.47$ & $4.50$
& $279$ & $4.17$ & $13.83$ & $4.23$ & $67.18$ & $13.83$
\\ \hline \hline 
\multirow{3}[6]{0.3cm}{\rotatebox[origin=c]{90}{{\hspace{0in}} {\sf Tx 1}}} &
$2.9$ & $2$ & $2.20$ & $1.25$ & $-$ & $-$ & $-$
& $38$ & $3.20$ & $7.87$ & $4.49$ & $22.25$ & $6.81$
\\ \cline{2-14}
&
$29$ & $1$ & $1.84$ & $0$ & $-$ & $-$ & $-$
& $34$ & $3.64$ & $10.27$ & $5.24$ & $37.68$ & $8.74$
\\ \cline{2-14}
&
$61$ & $2$ & $2.82$ & $1.26$ & $-$ & $-$ & $-$
& $40$ & $4.15$ & $18.06$ & $5.20$ & $52.13$ & $17.82$
\\ \hline \hline
\multirow{3}[6]{0.3cm}{\rotatebox[origin=c]{90}{{\hspace{0.2in}} {\sf Tx 2}}} &
$2.9$ & $4$ & $2.08$ & $3.96$ & $3.05$ & $28.12$ & $3.56$
& $33$ & $3.03$ & $4.81$ & $4.63$ & $19.71$ & $3.86$
\\ \cline{2-14}
&
$29$ & $5$ & $1.71$ & $5.83$ & $3.99$ & $30.61$ & $4.06$
& $39$ & $3.46$ & $6.23$ & $5.49$ & $33.40$ & $5.17$
\\ \hline \hline
\multirow{3}[6]{0.3cm}{\rotatebox[origin=c]{90}{{\hspace{0in}} {\sf Tx 3}}} &
$2.9$ & $11$ & $1.35$ & $2.95$ & $2.03$ & $48.04$ & $1.96$
& $34$ & $2.98$ & $6.22$ & $3.90$ & $28.11$ & $5.91$
\\ \cline{2-14}
&
$29$ & $11$ & $1.30$ & $1.84$ & $1.03$ & $65.11$ & $1.62$
& $33$ & $3.26$ & $7.41$ & $4.08$ & $49.61$ & $7.23$
\\ \cline{2-14}
&
$61$ & $17$ & $1.54$ & $3.88$ & $1.63$ & $66.94$ & $3.87$
& $239$ & $4.17$ & $13.03$ & $3.86$ & $73.00$ & $13.02$
\\ \hline
\hline
\multicolumn{14}{|c|}{${\bf  {\hspace{0.5in}}
Shopping \hspp Mall}$}
\\
\hline
&
& \multicolumn{6}{c||} { ${\sf LOS}$ } & \multicolumn{6}{c|} { ${\sf NLOS}$}
\\ \cline{1-14} 
&
$f_{\sf c}$ 
& ${\sf No. meas.}$
& ${\sf PLE}$
& $\sigma_{\sf CI}$ 
& $\alpha$
& $\beta'$
& $\sigma_{\sf ABG}$ 
& ${\sf No. meas.}$
& ${\sf PLE}$
& $\sigma_{\sf CI}$ 
& $\alpha$
& $\beta'$
& $\sigma_{\sf ABG}$ 
\\ \hline 
\multirow{3}[6]{0.3cm}{\rotatebox[origin=c]{90}{{\hspace{0in}} {\sf All}}} &
$2.9$ & $29$ & $1.93$ & $5.32$ & $1.74$ & $45.09$ & $5.29$
& $151$ & $2.61$ & $9.08$ & $2.81$ & $37.61$ & $9.07$
\\ \cline{2-14}
&
$29$ & $26$ & $1.98$ & $3.56$ & $1.62$ & $68.43$ & $3.45$
& $132$ & $2.76$ & $9.47$ & $2.96$ & $57.57$ & $9.45$
\\ \cline{2-14}
&
$61$ & $25$ & $2.05$ & $4.29$ & $1.90$ & $70.86$ & $4.27$
& $132$ & $2.98$ & $12.86$ & $2.27$ & $82.05$ & $12.70$
\\ \hline \hline 
\multirow{3}[6]{0.3cm}{\rotatebox[origin=c]{90}{{\hspace{0in}} {\sf Tx 1}}} &
$2.9$ & $12$ & $1.97$ & $7.04$ & $1.85$ & $43.70$ & $7.03$
& $82$ & $2.64$ & $9.85$ & $2.19$ & $50.31$ & $9.80$
\\ \cline{2-14}
&
$29$ & $8$ & $1.94$ & $1.67$ & $1.86$ & $63.06$ & $1.66$
& $61$ & $2.78$ & $10.87$ & $1.42$ & $87.44$ & $10.46$
\\ \cline{2-14}
&
$61$ & $8$ & $1.94$ & $1.30$ & $1.69$ & $72.40$ & $1.03$
& $58$ & $2.97$ & $12.53$ & $1.43$ & $97.28$ & $12.07$
\\ \hline \hline
\multirow{3}[6]{0.3cm}{\rotatebox[origin=c]{90}{{\hspace{0in}} {\sf Tx 2}}} &
$2.9$ & $8$ & $1.79$ & $3.04$ & $3.59$ & $5.75$ & $1.91$
& $38$ & $2.72$ & $5.89$ & $3.07$ & $34.29$ & $5.78$
\\ \cline{2-14}
&
$29$ & $9$ & $1.86$ & $2.96$ & $2.96$ & $39.69$ & $2.60$
& $44$ & $2.88$ & $7.38$ & $3.63$ & $45.60$ & $7.04$
\\ \cline{2-14}
&
$61$ & $8$ & $1.97$ & $5.06$ & $3.86$ & $30.51$ & $4.41$
& $42$ & $3.21$ & $11.72$ & $2.05$ & $92.46$ & $10.83$
\\ \hline \hline
\multirow{3}[6]{0.3cm}{\rotatebox[origin=c]{90}{{\hspace{0in}} {\sf Tx 3}}} &
$2.9$ & $9$ & $2.03$ & $3.33$ & $2.62$ & $31.07$ & $3.21$
& $31$ & $2.35$ & $8.61$ & $2.75$ & $33.96$ & $8.56$
\\ \cline{2-14}
&
$29$ & $9$ & $2.16$ & $3.19$ & $1.14$ & $79.86$ & $2.80$
& $27$ & $2.46$ & $6.65$ & $3.03$ & $50.86$ & $6.52$
\\ \cline{2-14}
&
$61$ & $9$ & $2.22$ & $3.93$ & $1.77$ & $76.03$ & $3.87$
& $32$ & $2.65$ & $12.46$ & $2.35$ & $73.89$ & $12.44$
\\ \hline
\ignore{
&
& \multicolumn{6}{c||} { ${\sf LOS}$ } & \multicolumn{6}{c|} { ${\sf NLOS}$}
\\ \cline{1-14} 
&
$f_{\sf c}$ 
& ${\sf No. meas.}$
& ${\sf PLE}$
& $X_{\sf CI}$ 
& $\alpha$
& $\beta$
& $X_{\sf ABG}$ 
& ${\sf No. meas.}$
& ${\sf PLE}$
& $X_{\sf CI}$ 
& $\alpha$
& $\beta$
& $X_{\sf ABG}$ 
\\ \hline 
\multirow{3}[6]{0.3cm}{\rotatebox[origin=c]{90}{{\hspace{0in}} {\sf All}}} &
$2.9$ & $6$ & $2.18$ & $4.41$ & $3.23$ & $-22.82$ & $3.35$
& $35$ & $2.95$ & $7.82$ & $4.32$ & $-29.19$ & $7.60$
\\ \cline{2-14}
&
$29$ & $7$ & $2.19$ & $4.37$ & $3.11$ & $-19.38$ & $3.47$
& $41$ & $3.07$ & $8.16$ & $4.40$ & $-28.30$ & $7.97$
\\ \cline{2-14}
&
$61$ & $6$ & $2.22$ & $4.84$ & $3.12$ & $-19.24$ & $4.19$
& $40$ & $3.27$ & $10.70$ & $5.18$ & $-40.59$ & $10.41$
\\ \hline 
}
\end{tabular}
\end{center}
\end{table*}

\begin{table*}[htb!]
\caption{Path Loss Coefficients for the Different Outdoor Settings.
$f_{\sf c}$ is in GHz, $\sigma_{\sf CI}, \sigma_{\sf ABG}, \beta'$ are in ${\rm dB}$.} 
\label{table1_outdoor}
\begin{center}
\begin{tabular}{|l
|c|| c|c|c|c|c|c|| c|c|c|c|c|c|}
\hline
\multicolumn{14}{|c|}{${\bf  {\hspace{0.5in}}
UMi, \hspp Street \hspp Canyon}$}
\\
\hline
&
& \multicolumn{6}{c||} { ${\sf LOS}$ } & \multicolumn{6}{c|} { ${\sf NLOS}$}
\\ \cline{1-14} 
&
$f_{\sf c}$ 
& ${\sf No. meas.}$
& ${\sf PLE}$
& $\sigma_{\sf CI}$ 
& $\alpha$
& $\beta'$
& $\sigma_{\sf ABG}$ 
& ${\sf No. meas.}$
& ${\sf PLE}$
& $\sigma_{\sf CI}$ 
& $\alpha$
& $\beta'$
& $\sigma_{\sf ABG}$ 
\\ \hline 
\multirow{3}[6]{0.3cm}{\rotatebox[origin=c]{90}{{\hspace{0in}} {\sf All}}} &
$2.9$ & $6$ & $2.18$ & $4.41$ & $3.23$ & $18.87$ & $3.35$
& $35$ & $2.95$ & $7.82$ & $4.32$ & $12.50$ & $7.60$
\\ \cline{2-14}
&
$29$ & $7$ & $2.19$ & $4.37$ & $3.11$ & $42.31$ & $3.47$
& $41$ & $3.07$ & $8.16$ & $4.40$ & $33.39$ & $7.97$
\\ \cline{2-14}
&
$61$ & $6$ & $2.22$ & $4.84$ & $3.12$ & $48.91$ & $4.19$
& $40$ & $3.27$ & $10.70$ & $5.18$ & $27.56$ & $10.41$
\\ \hline 
\hline
\multicolumn{14}{|c|}{${\bf  {\hspace{0.5in}}
Outside \hspp Qualcomm \hspp Building, \hspp Open \hspp Areas}$}
\\
\hline
&
& \multicolumn{6}{c||} { ${\sf LOS}$ } & \multicolumn{6}{c|} { ${\sf NLOS}$}
\\ \cline{1-14} 
&
$f_{\sf c}$ 
& ${\sf No. meas.}$
& ${\sf PLE}$
& $\sigma_{\sf CI}$ 
& $\alpha$
& $\beta'$
& $\sigma_{\sf ABG}$ 
& ${\sf No. meas.}$
& ${\sf PLE}$
& $\sigma_{\sf CI}$ 
& $\alpha$
& $\beta'$
& $\sigma_{\sf ABG}$ 
\\ \hline 
\multirow{3}[6]{0.3cm}{\rotatebox[origin=c]{90}{{\hspace{0in}} {\sf All}}} &
$2.9$ & $47$ & $2.41$ & $4.60$ & $3.03$ & $28.54$ & $4.56$
& $13$ & $3.01$ & $4.00$ & $5.91$ & $-21.29$ & $3.07$
\\ \cline{2-14}
&
$29$ & $49$ & $2.73$ & $5.73$ & $2.46$ & $67.31$ & $5.72$
& $13$ & $3.39$ & $8.03$ & $8.70$ & $-53.36$ & $6.53$
\\ \cline{2-14}
&
$61$ & $37$ & $2.83$ & $6.78$ & $5.40$ & $13.38$ & $6.24$
& $6$ & $3.42$ & $1.97$ & $0.08$ & $137.81$ & $0.83$
\\ \hline 
\hline
\multicolumn{14}{|c|}{${\bf  {\hspace{0.5in}}
Outside \hspp Qualcomm \hspp Building, \hspp Parking \hspp Structures}$}
\\
\hline
&
& \multicolumn{6}{c||} { ${\sf LOS}$ } & \multicolumn{6}{c|} { ${\sf NLOS}$}
\\ \cline{1-14} 
&
$f_{\sf c}$ 
& ${\sf No. meas.}$
& ${\sf PLE}$
& $\sigma_{\sf CI}$ 
& $\alpha$
& $\beta'$
& $\sigma_{\sf ABG}$ 
& ${\sf No. meas.}$
& ${\sf PLE}$
& $\sigma_{\sf CI}$ 
& $\alpha$
& $\beta'$
& $\sigma_{\sf ABG}$ 
\\ \hline 
\multirow{3}[6]{0.3cm}{\rotatebox[origin=c]{90}{{\hspace{0.2in}} {\sf All}}} &
$2.9$ & $10$ & $2.82$ & $13.54$ & $0.82$ & $83.95$ & $8.26$
& $35$ & $3.23$ & $8.54$ & $2.85$ & $49.94$ & $8.44$
\\ \cline{2-14}
&
$29$ & $9$ & $2.94$ & $21.02$ & $-0.49$ & $132.71$ & $9.57$
& $36$ & $3.44$ & $10.50$ & $2.21$ & $88.41$ & $9.63$
\\ \hline 

\end{tabular}
\end{center}
\end{table*}

\begin{figure*}[htb!]
\begin{center}
\begin{tabular}{cc}
\includegraphics[height=2.3in,width=3.15in] {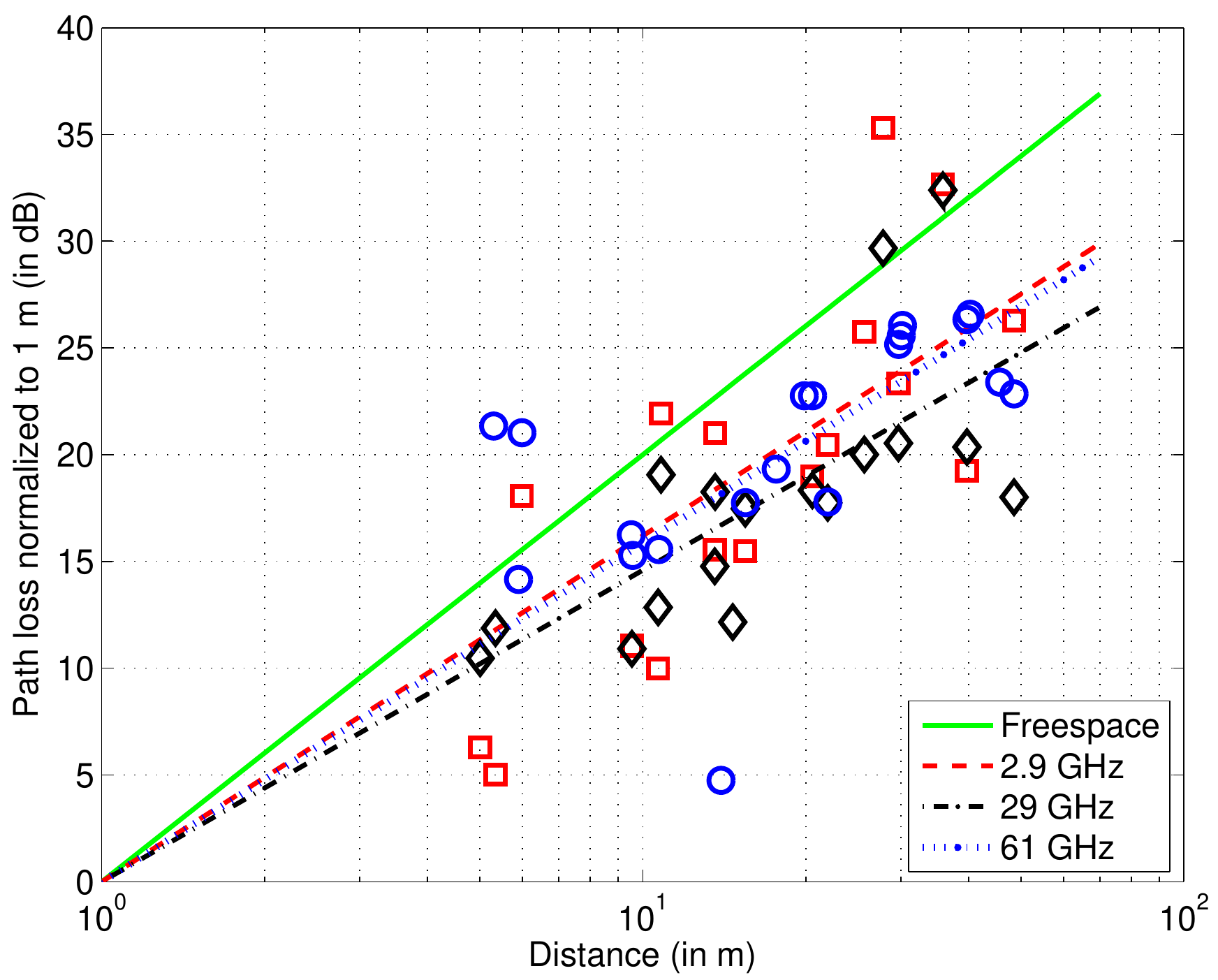}
&
\includegraphics[height=2.3in,width=3.15in] {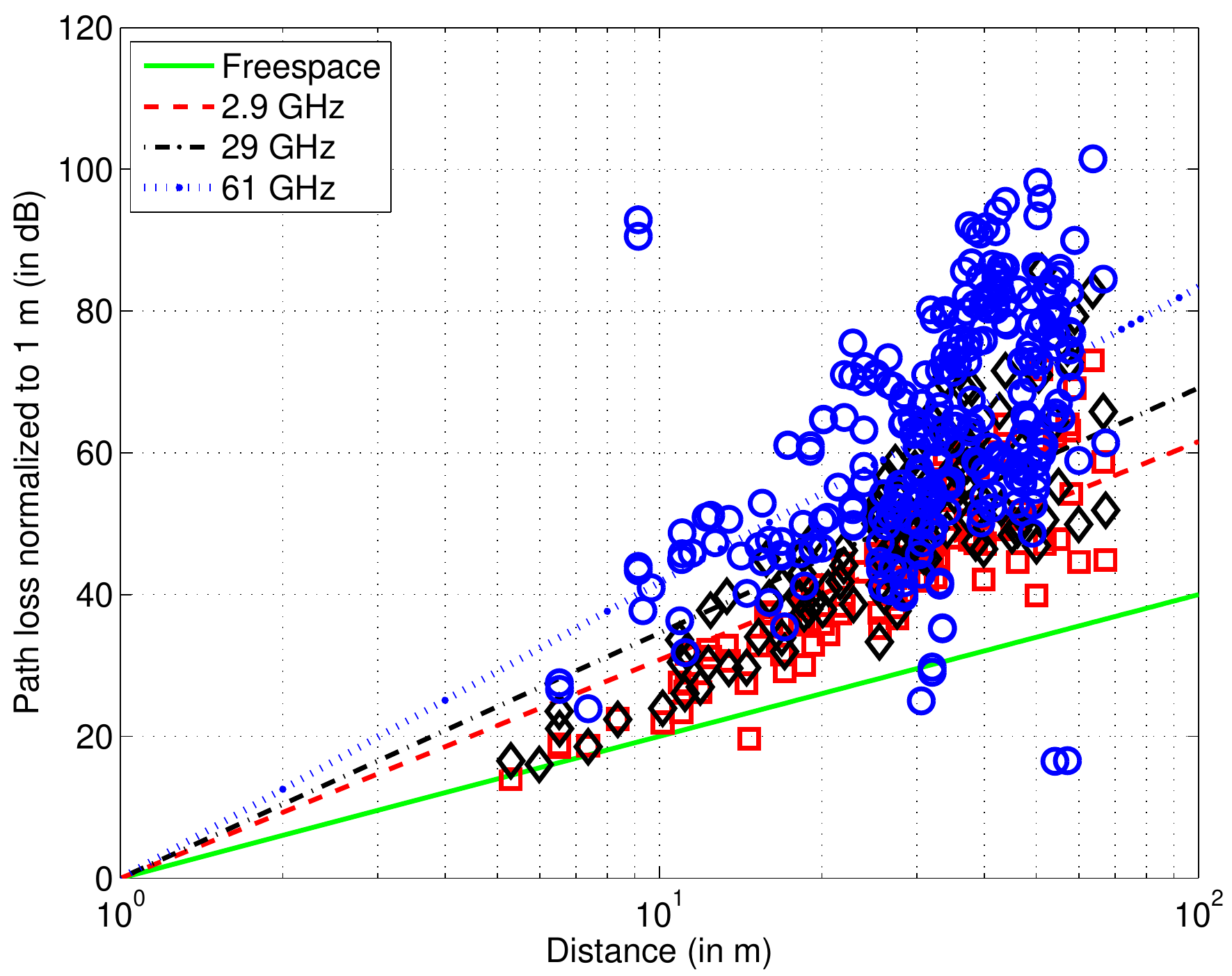}
\\ {\hspace{0.35in}} (a) & {\hspace{0.12in}} (b)
\\
\includegraphics[height=2.3in,width=3.15in]
{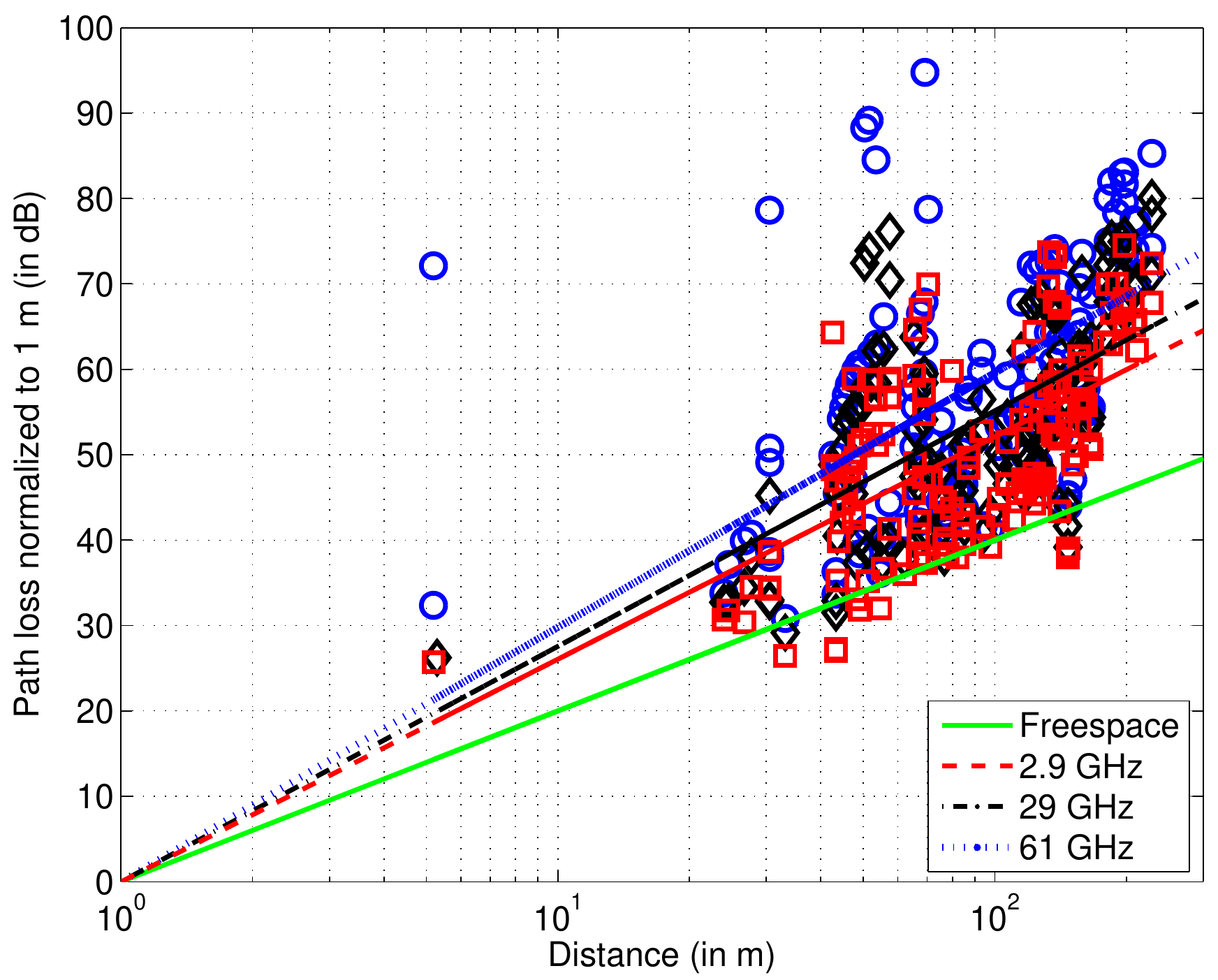}
&
\includegraphics[height=2.3in,width=3.15in]
{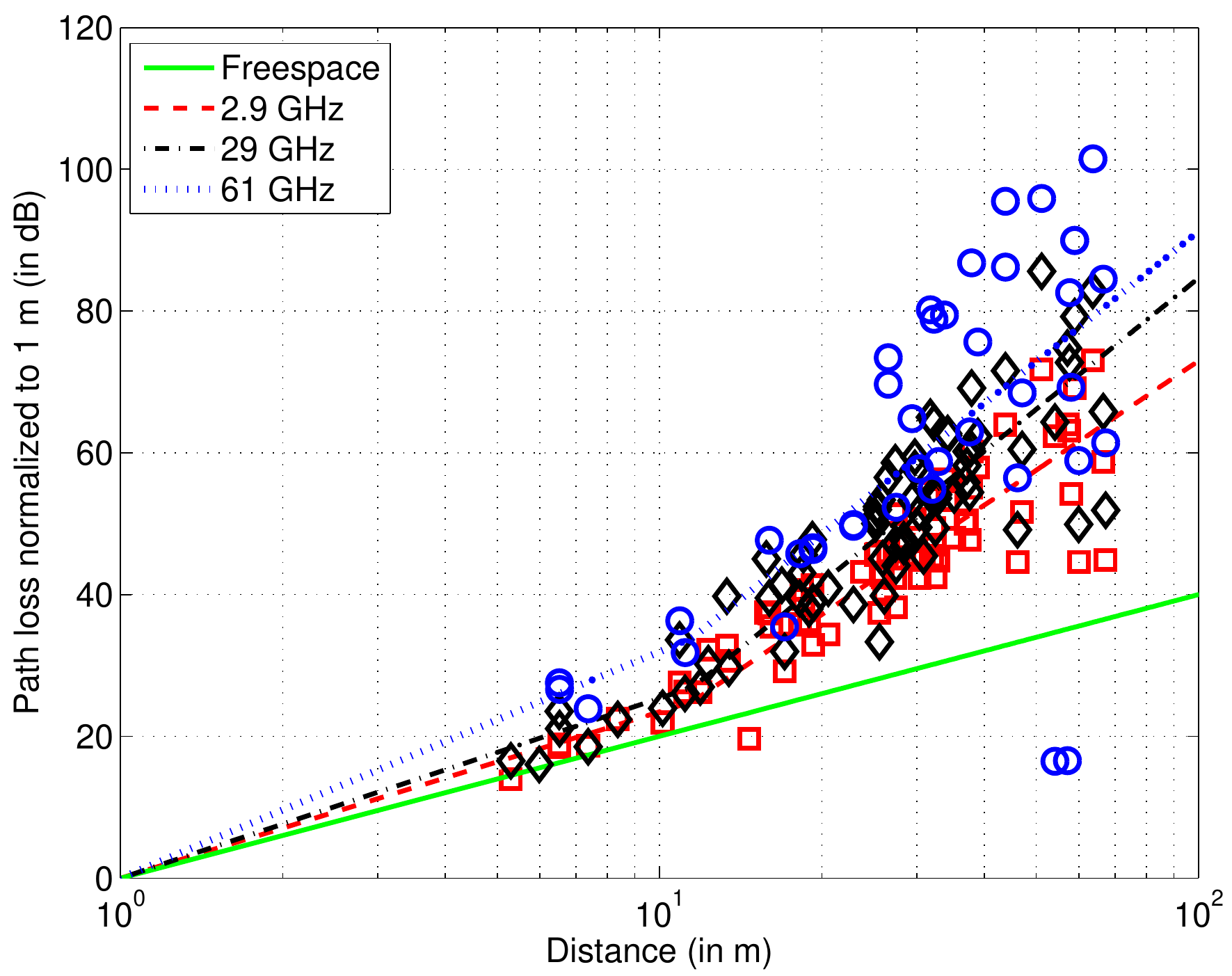}
\\ {\hspace{0.35in}} (c) & {\hspace{0.12in}} (d)
\end{tabular}
\caption{\label{fig_PLfits}
Path loss fits with the CI model for (a) LOS and (b) NLOS links for indoor office,
and (c) with NLOS links for shopping mall settings. (d) Dual-slope model for NLOS links
in third floor of indoor office.
}
\end{center}
\vspace{-5mm}
\end{figure*}

\subsection{Path Loss}
\label{sec3a}
The total received power from omni-directional antenna measurements is used to
estimate the path loss model for $2.9$, $29$ and $61$ GHz. The path loss from
measurements is fitted with two popular frequency-dependent models in the
literature~\cite{5G_whitepaper}: the close-in (CI) reference model
and the Alpha-Beta-Gamma (ABG) model. In the CI model, the path loss at a
distance of $d$ m is given as
\begin{eqnarray}
{\sf PL}(d) 
= {\sf PL}(d_0) 
+ {\sf PLE} \cdot 10 \log_{10}(d/d_0) + X_{\sf CI} \hsp \hsp [{\sf in \hspp dB}]
\label{eq_pathloss}
\end{eqnarray}
where ${\sf PL}(d_0)$ 
is the path loss at a reference distance of $d_0$ m, ${\sf PLE}$ is the
path loss exponent (PLE), and $X_{\sf CI} \sim {\cal N}(0, \sigma_{\sf CI}^2)$ models
log-normal shadowing. With the reference distance typically set to $d_0 = 1$ m, the
reference path loss ${\sf PL}(d_0) 
= 20 \cdot \log_{10} \left( \frac{4 \pi d_0}{\lambda} \right)$
is removed from the measurement data to normalize the path loss to $0$ dB at
$d_0$ for all the three frequencies thus allowing a direct comparison. An estimate
of the PLE and $\sigma_{\sf CI}$ are obtained through a least-squares fit of the
parameters to the measurement data. 

In the ABG model, the path loss\footnote{In scenarios where the ABG model is fitted
across a single frequency, $\beta$ and $\gamma$ can be combined to lead to a simplified
parameter $\beta'$ as in~(\ref{eq_pathloss_ABG}).} is given as
\begin{eqnarray}
{\sf PL}(d) 
& = & 10 \alpha \cdot \log_{10} \left( \frac{d}{d_0} \right) +
\underbrace{ \beta
+ 10 \gamma \cdot \log_{10} \left( \frac{f_{\sf c} } { 1 \hsppp {\sf GHz} } \right)
}_{ = \beta'}
\nonumber \\
& & {\hspace{0.2in}} + X_{\sf ABG} \hsp \hsp [{\sf in \hspp dB}]
\label{eq_pathloss_ABG}
\end{eqnarray}
where $\alpha$ and $\gamma$ capture how the path loss changes with distance
and frequency, $\beta$ is an optimized parameter and $X_{\sf ABG}
\sim {\cal N}(0, \sigma_{\sf ABG}^2)$ models log-normal shadowing. The CI and ABG models
trade off\footnote{A better fit can be expected with the ABG model since the two
parameter CI framework can be subsumed within the four parameter ABG framework. Whether
the increase in number of parameters results in a substantially better model fit is a
question of further interest, answered subsequently.} explanatory power at the
cost of more model parameters (better fit with
the four parameter ABG model). As in the CI case, the model parameters in the ABG
case are also learned using a least-squares fit of the parameters to measurement data.
Tables~\ref{table1_indoor} and~\ref{table1_outdoor} present the
parameters\footnote{ABG model parameters are not presented in scenarios with few
data points.} for the CI and ABG models in both LOS and NLOS settings at different
carrier frequencies and measurement scenarios.
Table~\ref{table1_indoor} focusses on the indoor scenario
(office and mall), whereas Table~\ref{table1_outdoor} focusses on the outdoor scenario
(downtown New Brunswick and outside Qualcomm building). Figs.~\ref{fig_PLfits}(a)-(c)
present the path loss fits with the CI model in the office and mall settings.

In the indoor office setting, both third and fourth floor data were combined together
in obtaining a global estimate of PLE and shadowing factors with both models across the
building. The best
fit PLEs and shadowing factors for NLOS links at $2.9$, $29$ and $61$ GHz are
$3.1$, $3.5$ and $4.2$, and $6.6$, $8.3$ and $13.8$ dB, respectively.
On the other hand, path loss fits conditioned on third floor locations alone suggest a
better fit with a dual-slope model corresponding to a breakpoint distance of ${d}_{\sf BP}$
than a single-slope model:
\begin{align}
& {\sf PL}(d) 
- {\sf PL}(d_0) 
\nonumber \\
&
=
\left\{ \begin{array}{l}
{\sf PLE}_1 \cdot 10 \log_{10}(d/d_0) + X_{\sf CI}^1
\hspp {\sf if} \hspp
d < {d}_{\sf BP} \\
{\sf PLE}_2 \cdot 10 \log_{10}(d/ {d}_{\sf BP} ) +
{\sf PLE}_1 \cdot 10 \log_{10}( {d}_{\sf BP} ) +
X_{\sf CI}^2
\\ {\hspace{2.4in}}
\hspp {\sf if} \hspp
d \geq {d}_{\sf BP}
\end{array}
\right.
\end{align}
For example, at $2.9$ GHz, we obtain $d_{\sf BP} = 11.5$ m,
${\sf PLE}_1 = 2.35$, ${\sf PLE}_2 = 5.12$, $\sigma_{\sf CI}^1 = 2.03$ dB,
$\sigma_{\sf CI}^2 = 5.98$ dB leading to a net shadowing factor of $5.68$ dB. On the
other hand, the single slope model results in ${\sf PLE} = 3.13$ and $\sigma_{\sf CI}
= 6.69$ dB. The path loss fits in the NLOS setting at different frequencies are
presented in Fig.~\ref{fig_PLfits}(d). 
These observations suggest
that two distinct modes of communications may be possible in indoor settings (long walkways
and office rooms in fourth floor vs.\ primarily cubicles and conference rooms in third
floor): predominantly reflected and diffracted paths at $d < d_{\sf BP}$ and
$d \geq d_{\sf BP}$, respectively.
PLEs for LOS links in the indoor office setting are considerably lower:
$1.6$, $1.5$ and $1.6$ at $2.9$, $29$ and $61$ GHz, respectively. The discrepancy of
lower PLE at $29$ GHz is ascribed to waveguide effects in the indoor office setting and/or
changes in material properties at higher frequencies. The $\beta'$ parameter estimated with
the ABG model shows wide variations, also documented in other works such
as~\cite{shu_sun_vtc2016}.

In the mall use-case with NLOS links, the PLEs and $\sigma_{\sf CI}$ at these three frequencies
are $2.6$, $2.8$ and $3.0$, and $9.1$, $9.5$ and $12.9$ dB, respectively. In the LOS link
case, the PLEs and shadowing factors are considerably lower: $1.9$, $2.0$ and $2.1$, and
$5.3$, $3.6$ and $4.3$ dB, respectively. Based on a more detailed study of parameters
from individual transmitter locations, we have the following broad conclusions which is
also in agreement with the main conclusions
of~\cite{5G_whitepaper,shu_sun_vtc2016,shu_sun2016}: i) A
general increase in the PLE (especially NLOS links) with frequency, and ii) Better
propagation for the LOS link over NLOS links. Log-normal shadowing studies
suggest its general increase with frequency and distance as well as the utility of a
piecewise linear model (different models valid above and below a breakpoint distance)
in the NLOS setting.

Table~\ref{table1_outdoor} provides a summary of the path loss parameters in the
outdoor settings. The main conclusions from these studies are: i) Consistent increase
in the PLE in both LOS and NLOS cases in all the scenarios, and ii) While the shadow
fading parameters generally increase with frequency, inconsistent trends are occasionally
seen at higher carrier frequencies due to radar cross-section\footnote{Radar cross-section
tells us how much more reflected energy is received when compared to reflection from a
ball having a cross-section of $1$ sq m, or equivalently how much bigger a ball is
needed to have the same effect.} effect of certain reflectors. From a performance
comparison between the CI and ABG models, in all the settings considered here, we
observe that $\sigma_{\sf CI}$ is comparable with $\sigma_{\sf ABG}$
provided that there are enough measurements to ensure parameter consistency. Thus, the
CI model appears to provide a comparable fit relative to the ABG model with a smaller
number of parameters and is hence {\em preferable.} Similar conclusions have also
been made in~\cite{shu_sun_vtc2016,shu_sun2016} from more general parameter stability considerations.


\begin{figure*}[htb!]
\begin{center}
\begin{tabular}{cc}
\includegraphics[height=2.3in,width=3.15in] {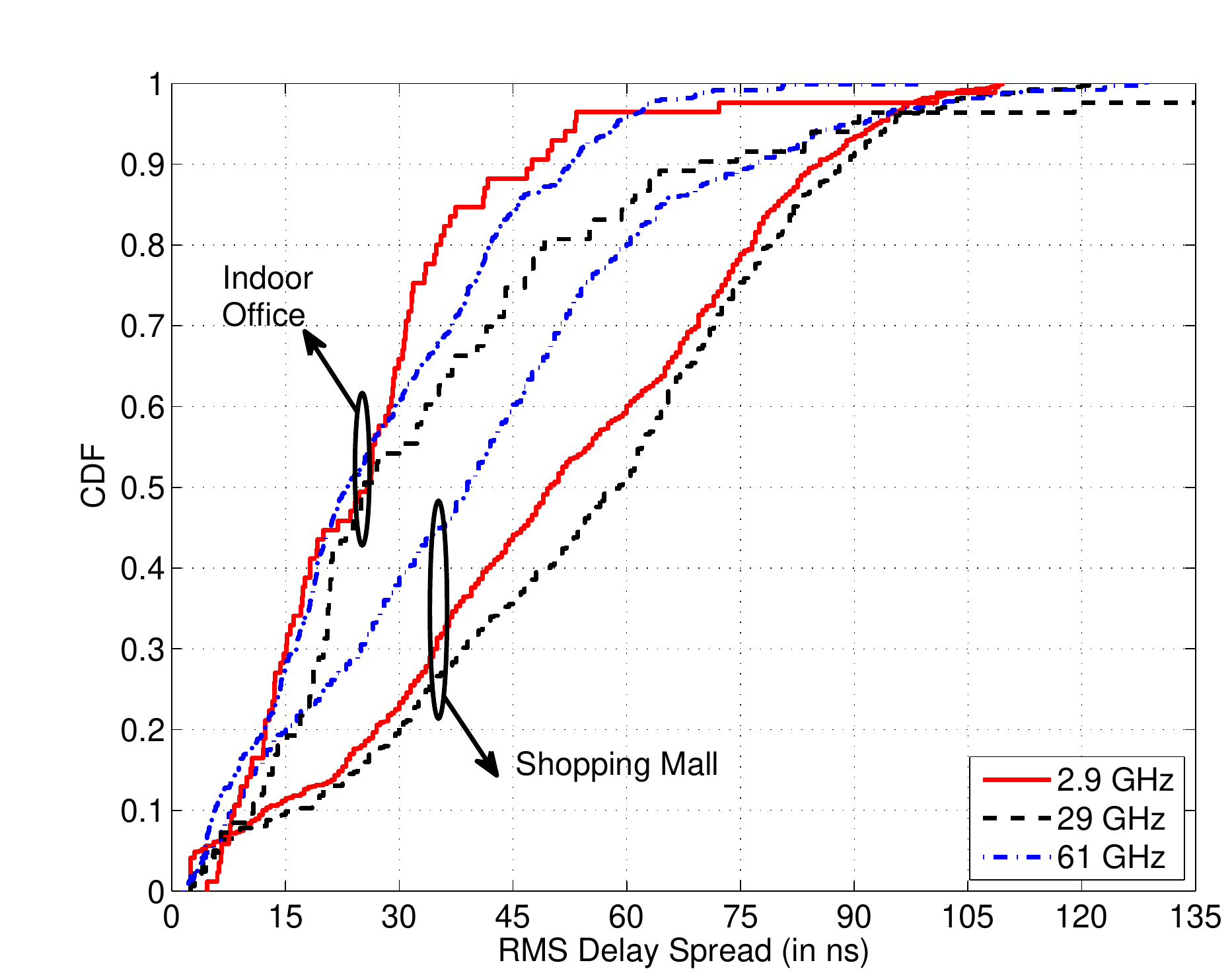}
&
\includegraphics[height=2.3in,width=3.15in] {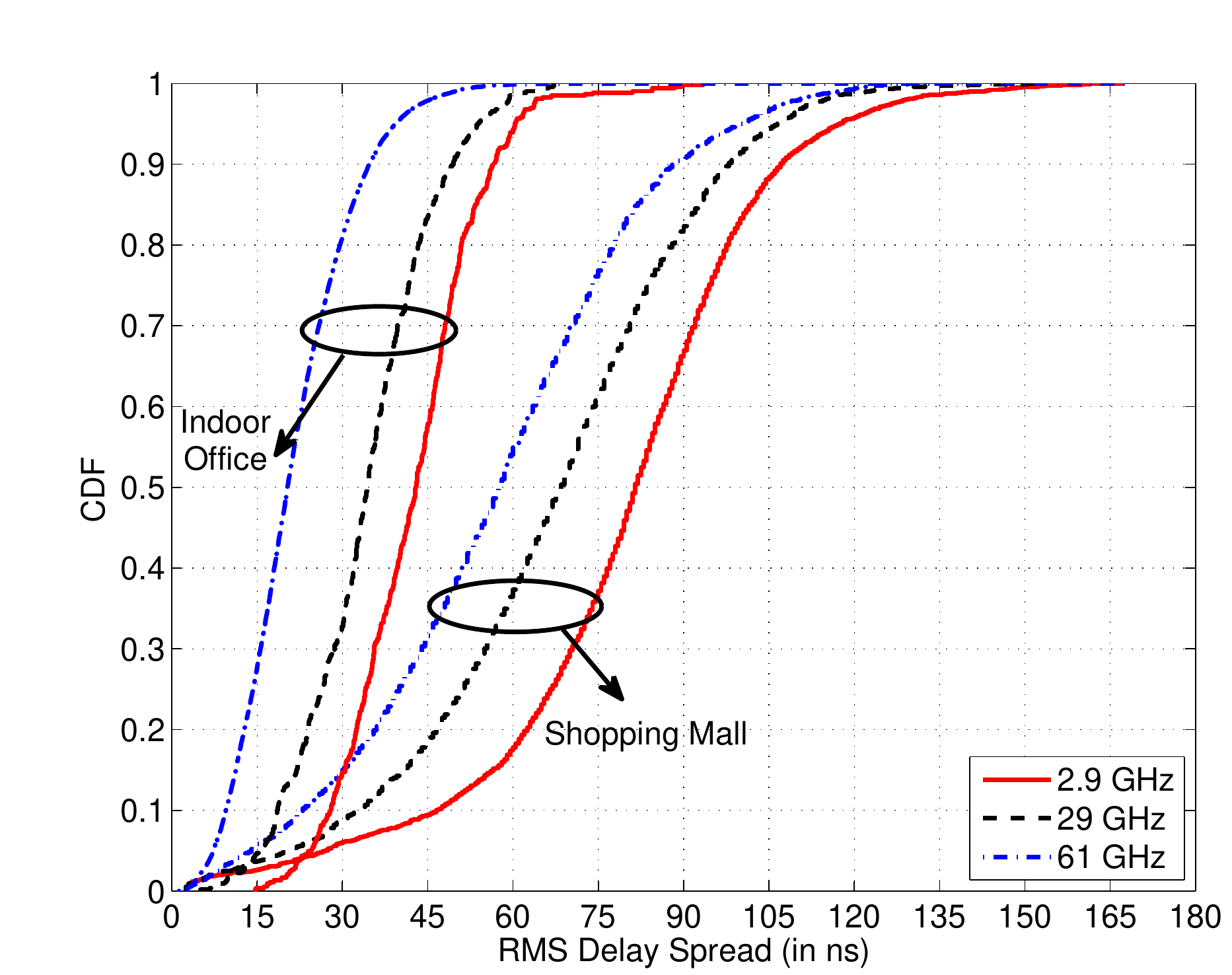}
\\ {\hspace{0.35in}} (a) & {\hspace{0.12in}} (b)
\\
\includegraphics[height=2.3in,width=3.15in]
{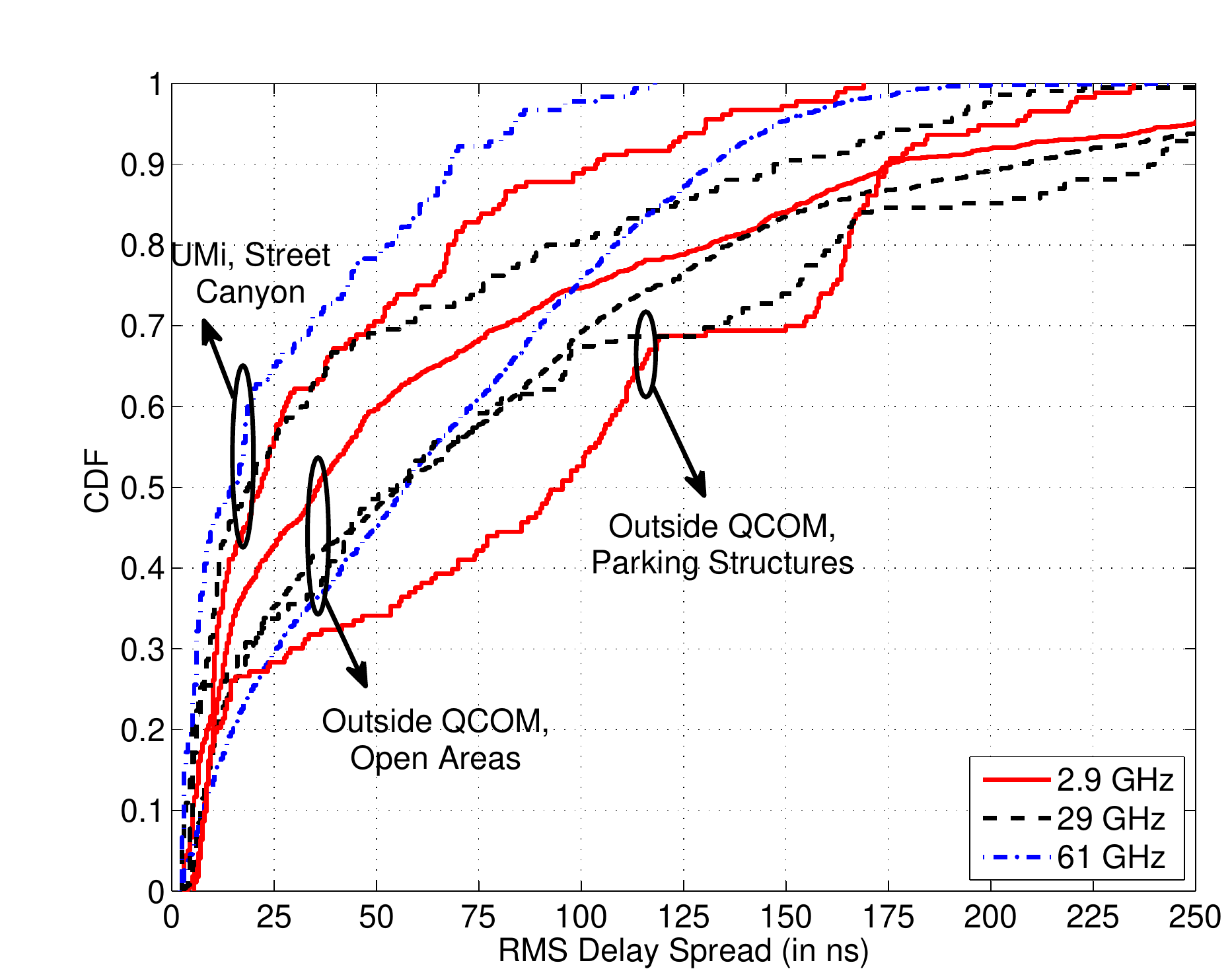}
&
\includegraphics[height=2.3in,width=3.15in] {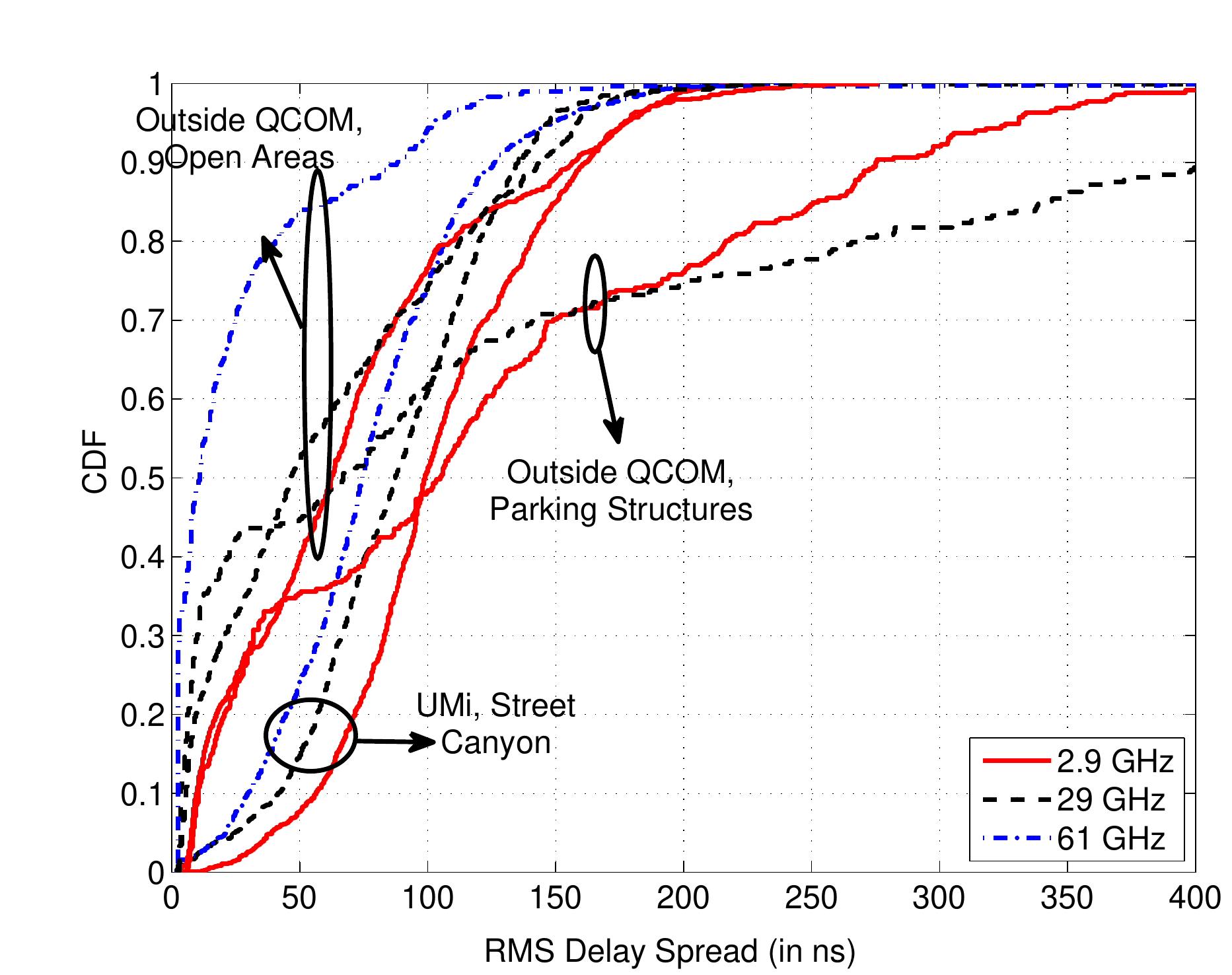}
\\ {\hspace{0.35in}} (c) & {\hspace{0.12in}} (d)
\end{tabular}
\caption{\label{fig_rms_delay_spread}
CDF of RMS delay spread in (a)-(b) indoor LOS and NLOS, and (c)-(d)
outdoor LOS and NLOS settings.
}
\end{center}
\vspace{-5mm}
\end{figure*}

\subsection{Delay Spread}
\label{sec3b}
The excess delay (denoted as $\tau_{\sf excess}$) and RMS delay spread (denoted
as $\tau_{\sf rms}$) with omni-directional scans across 
different environments are studied. If $\tau_i$ and $p_i$ denote the delay and power
corresponding to the $i$-th tap in a certain omni-directional scan, the excess delay
and RMS delay spread are computed as:
\begin{eqnarray}
\tau_{\sf excess} & = & \frac{ \sum_i \tau_i p_i }{ \sum_i p_i }
\\
\tau_{\sf rms} & = & \sqrt{ \frac{ \sum_i \tau_i^2 p_i }{ \sum_i p_i} -
\left( \frac{ \sum_i \tau_i p_i }{ \sum_i p_i } \right)^2 }. 
\end{eqnarray}

In the indoor office setting, the longest end-to-end
delay is $250$ ns; any delay beyond this value is a result of reflections. For excess
delay, an exponential distribution to the data for each link type and frequency band
is fitted. The means of the exponential at $2.9$, $29$ and $61$ GHz for the excess delay
of the combined third and fourth floor measurements with NLOS links are given by
$\lambda^{-1} = 93.4, 82.3$ and $52.2$ ns, respectively. This trend is as expected
given the difference in propagation characteristics at higher frequencies.
The CDF of RMS delay spreads and the parameters associated with an exponential fit for
NLOS links are provided in Fig.~\ref{fig_rms_delay_spread} and Table~\ref{table2},
respectively. The corresponding numbers for the exponential fit to the excess delay
at $2.9$, $29$ and $61$ GHz in the LOS case are $\lambda^{-1} = 65.8$, $71.9$ and
$33.3$ ns. For LOS links, the mean of the excess delay is actually higher at $29$ GHz.
The RMS delay spread for LOS links 
illustrates this difference through a heavier tail at larger delay values.

The parameters associated with the RMS delay spread across all transmitter and receiver
locations with omni-directional antenna scans at $2.9$, $29$ and $61$ GHz in the
shopping mall are also presented in 
Table~\ref{table2} for NLOS and LOS links. 
As in the indoor office setting, an increase in frequency reduces the RMS delay spread
for NLOS links and the RMS delay spread for LOS links at $29$ GHz is in general larger
than at $2.9$ GHz. Similar behavior is seen for the excess delay distributions.
Table~\ref{table2} also presents 
the RMS delay spread parameters for different scenarios in the outdoor case. In general,
the delay spread in the outdoor setting is larger than in the indoor setting with some
tail values corresponding to strong but significantly delayed sub-dominant clusters/paths.

The main conclusions from our studies are: i) Delay spread for NLOS links generally
decrease with increase in frequency. ii) Delay spread for LOS links decrease with
frequency in dense environments. iii) Delay spread for LOS links in non-dense
environments shows an inconsistent behavior as frequency increases. While the
literature does not offer a good explanation for this inconsistent behavior, the
plausible explanations for it are:
\begin{itemize}
\item {\em Waveguide effect} where long enclosures such as walkways/corridors,
dropped/false ceilings, etc., tend to propagate electromagnetic energy via
alternate modes/more reflective paths decreasing the PLE (often even below the
free-space PLE of $2$) and increasing the delay spread as frequency increases.

\item {\em Radar cross-section effect} where seemingly small objects that do not
participate in electromagnetic propagation at lower frequencies show up at higher
frequencies. Such behavior happens as the wavelength approaches the roughness of
surfaces (e.g., walls, light poles, etc.).
\end{itemize}

Since millimeter wave systems are likely to be used with beamforming,
it is of interest in understanding the beamformed delay spread of the channel relative
to that with an omni-directional scan. In this context, we note that in general,
the beamformed delay spread is smaller than the omni delay spread. However, for most
scenarios of interest in the indoor setting, this reduction is only by a small amount.
A simple explanation for this observation is that indoor millimeter wave channels are sparse
with few dominant clusters/paths. On the other hand, in the outdoor setting, the beamformed
delay spread for the tail values can be significantly smaller than the omni delay spread.
Thus, the effect of the significantly delayed sub-dominant clusters/paths get mitigated
with beamforming.

\begin{table*}[htb!]
\caption{Statistics of RMS Delay Spread}
\label{table2}
\begin{center}
\begin{tabular}{
|c|c|  |c|c||c|c||c|c|}
\hline
${\sf Metric}$ &
${\sf Parameter}$ & \multicolumn{2}{c||}{ ${\sf f_c} = 2.9 \hspp {\sf GHz}$}
& \multicolumn{2}{c||} {${\sf f_c} = 29 \hspp {\sf GHz}$}
& \multicolumn{2}{c|}{${\sf f_c} = 61 \hspp {\sf GHz}$}
\\ \hline \hline
${\sf \bf Indoor \hspp Office}$
& & ${\sf LOS}$ & ${\sf NLOS}$ & ${\sf LOS}$ & ${\sf NLOS}$ & ${\sf LOS}$ & ${\sf NLOS}$
\\ \hline
${\sf Delay \hspp spread}$ & ${\sf Median}$ $({\sf in \hsppp ns})$ & $25.72$ & $42.89$ &
$25.39$ & $34.34$ & $23.10$ & $20.36$ \\ \hline
$\log_{10}( {\sf Delay \hspp spread})$  & ${\sf Mean}$ & $-7.67$ & $-7.39$
& $-7.56$ & $-7.49$ & $-7.68$ & $-7.72$ \\ \hline
& ${\sf Std.}$ & $0.28$ & $0.13$ & $0.39$ & $0.17$ & $0.35$ & $0.23$
\\ \hline \hline
${\sf \bf Shopping \hspp Mall}$
& & ${\sf LOS}$ & ${\sf NLOS}$ & ${\sf LOS}$ & ${\sf NLOS}$ & ${\sf LOS}$ & ${\sf NLOS}$
\\ \hline
${\sf Delay \hspp spread}$ & ${\sf Median}$ $({\sf in \hsppp ns})$ & $50.0$ & $81.5$ &
$59.0$ & $68.5$ & $39.0$ & $57.5$ \\ \hline
$\log_{10}( {\sf Delay \hspp spread})$  & ${\sf Mean}$ & $-7.40$ & $-7.15$
& $-7.35$ & $-7.23$ & $-7.52$ & $-7.31$ \\ \hline
& ${\sf Std.}$ & $0.38$ & $0.25$ & $0.35$ & $0.26$ & $0.38$ & $0.27$
\\ \hline \hline
${\sf \bf UMi, \hspp Street \hspp Canyon}$
& & ${\sf LOS}$ & ${\sf NLOS}$ & ${\sf LOS}$ & ${\sf NLOS}$ & ${\sf LOS}$ & ${\sf NLOS}$
\\ \hline
${\sf Delay \hspp spread}$ & ${\sf Median}$ $({\sf in \hsppp ns})$ & $21.75$
& $99.0$ & $18.75$ & $87.25$ & $14.75$ & $74.5$ \\ \hline
$\log_{10}( {\sf Delay \hspp spread})$  & ${\sf Mean}$ & $-7.65$ & $-7.02$ & $-7.67$ &
$-7.11$ & $-7.85$ & $-7.18$ \\ \hline
& ${\sf Std.}$ & $0.48$ & $0.20$ & $0.59$ & $0.28$ & $0.51$ & $0.30$
\\ \hline \hline
${\sf \bf Outside \hspp QCOM, \hspp Open \hspp Areas}$
& & ${\sf LOS}$ & ${\sf NLOS}$ & ${\sf LOS}$ & ${\sf NLOS}$ & ${\sf LOS}$ & ${\sf NLOS}$
\\ \hline
${\sf Delay \hspp spread}$ & ${\sf Median}$ $({\sf in \hsppp ns})$ & $35.5$ & $105.0$ &
$55.5$ & $67.0$ & $57.0$ & $11.0$ \\ \hline
$\log_{10}( {\sf Delay \hspp spread})$  & ${\sf Mean}$ & $-7.45$ & $-7.15$
& $-7.36$ & $-7.36$ & $-7.38$ & $-7.95$ \\ \hline
& ${\sf Std.}$ & $0.52$ & $0.52$ & $0.54$ & $0.75$ & $0.47$ & $0.57$
\\ \hline \hline
${\sf \bf Outside \hspp QCOM, \hspp Parking \hspp Structures}$
& & ${\sf LOS}$ & ${\sf NLOS}$ & ${\sf LOS}$ & ${\sf NLOS}$ & ${\sf LOS}$ & ${\sf NLOS}$
\\ \hline
${\sf Delay \hspp spread}$ & ${\sf Median}$ $({\sf in \hsppp ns})$ & $95.5$ & $62.5$ &
$55.0$ & $46.5$ & $-$ & $-$ \\ \hline
$\log_{10}( {\sf Delay \hspp spread})$  & ${\sf Mean}$ & $-7.26$ & $-7.31$
& $-7.38$ & $-7.44$ & $-$ & $-$ \\ \hline
& ${\sf Std.}$ & $0.52$ & $0.43$ & $0.62$ & $0.51$ & $-$ & $-$
\\ \hline 
\end{tabular}
\end{center}
\end{table*}

\section{Material Measurements}
\label{sec4}
Outdoor-to-indoor coverage critically depends on the reflection and penetration of mmW signals
through various materials found in residential/office buildings such as sheetrock,
concrete, glass, wood, etc.

\subsection{Reflection Response}
Towards this end, material measurements were performed
with a {\em completely synchronized} signal generator and signal analyzer sweeping the
$22$-$43$ GHz range. A horn antenna with rotational stages for easy adjustment of
polarization, a gain of $25$ dBi and $10^{\sf o}$ beamwidth was used in the studies.
The antenna was placed at about $1.5$-$2.5$ foot distance from the tested sample and
incidence angles are varied in the studies. Absorber panels were used to contain
reflections from the background objects surrounding the test site. A reference curve
was obtained by placing a ``perfect'' reflecting plate (a $2 \times 2$ sq ft aluminum
plate) and sweeping over the same frequency range to obtain the reflected energy.

\begin{figure*}[htb!]
\begin{center}
\begin{tabular}{cc}
\includegraphics[height=2.0in,width=3.1in] {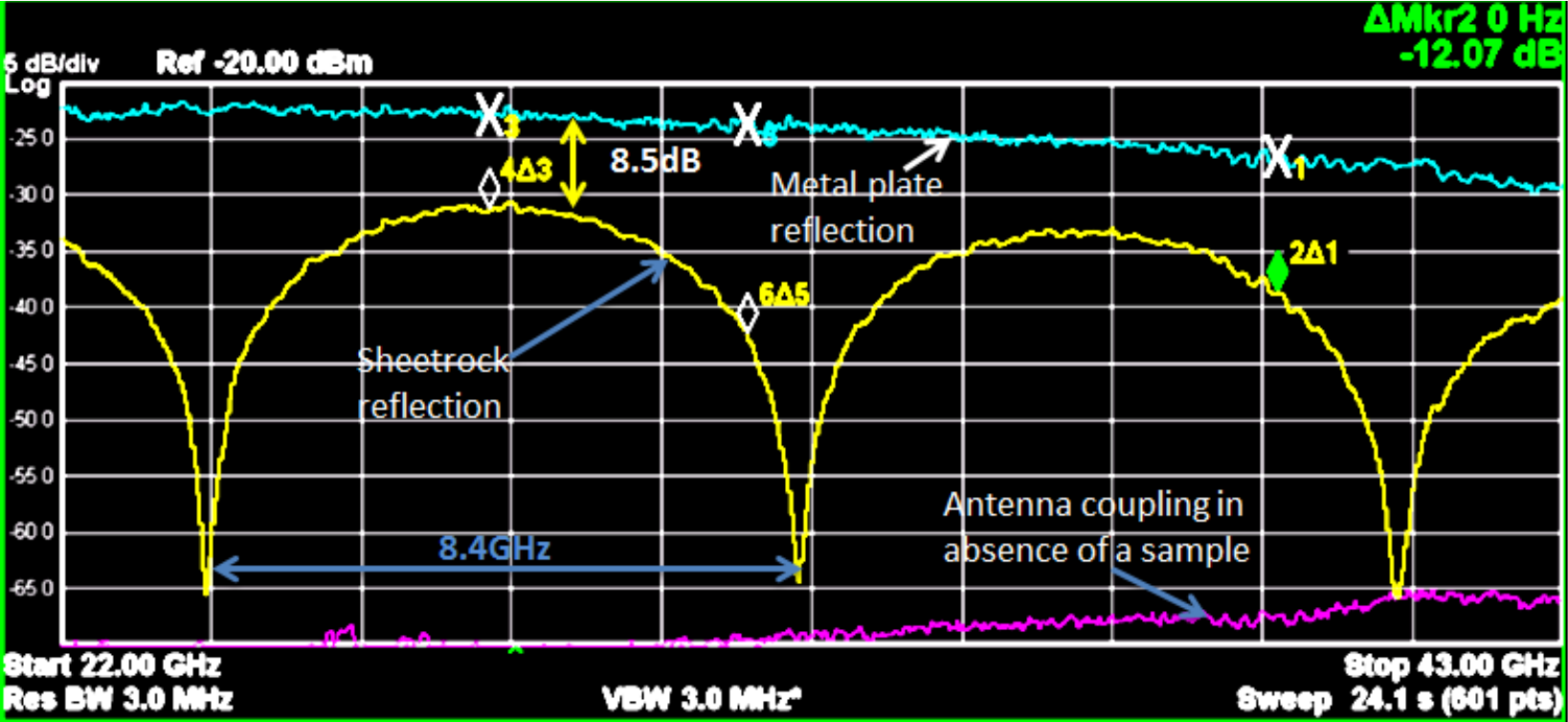}
&
\includegraphics[height=2.0in,width=3.1in] {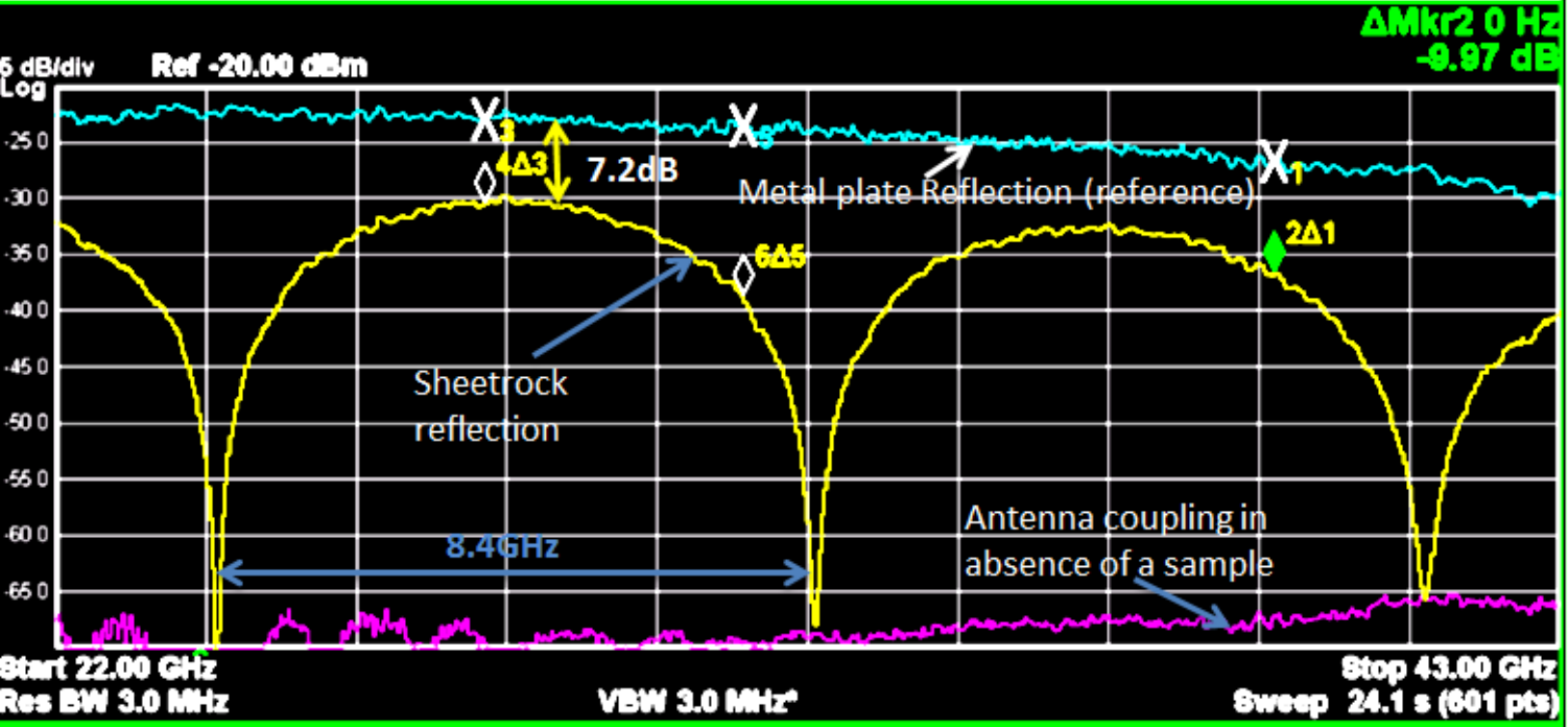}
\\
{\hspace{0.35in}} (a) & {\hspace{0.12in}} (b)
\\
\includegraphics[height=2.0in,width=3.1in] {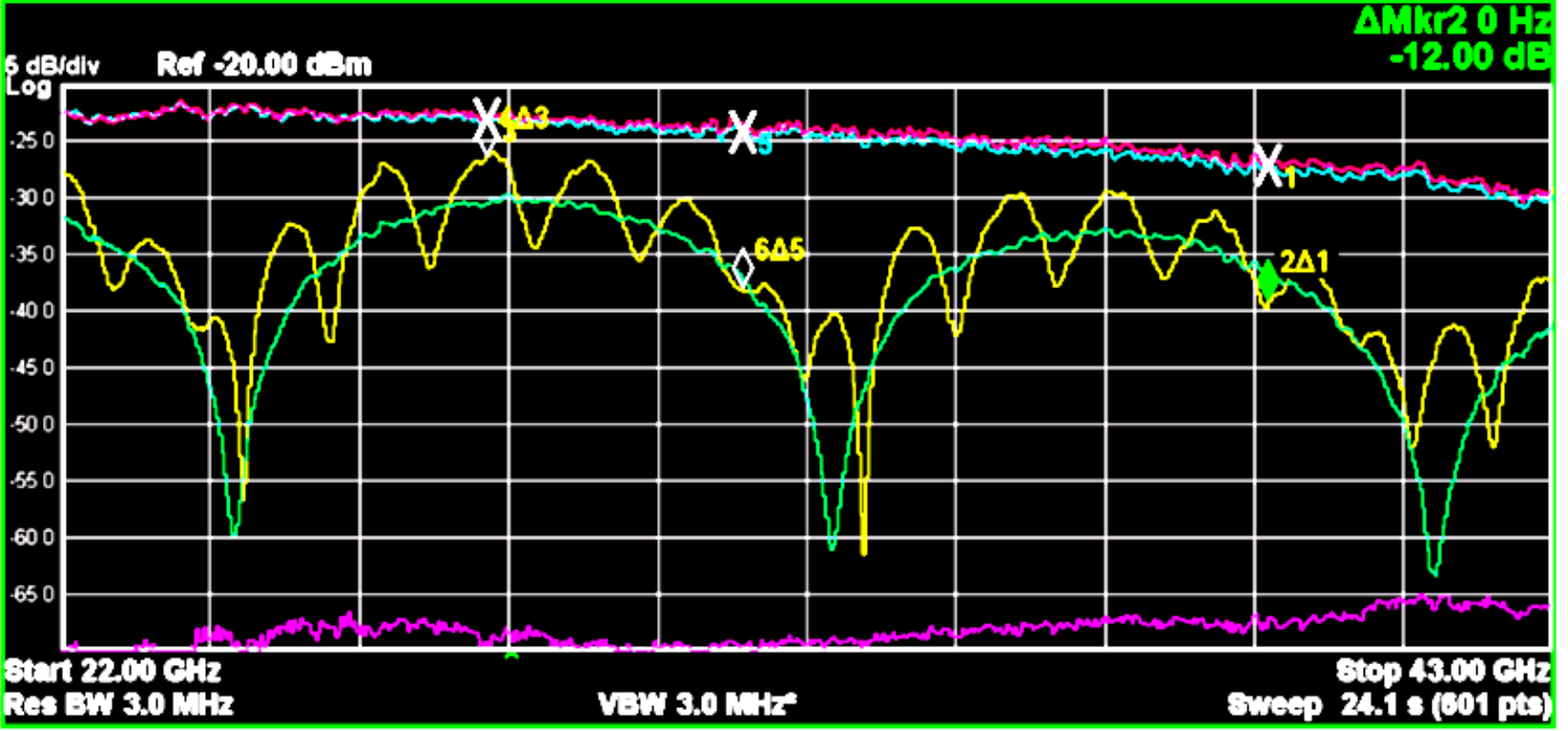}
&
\includegraphics[height=2.0in,width=3.1in] {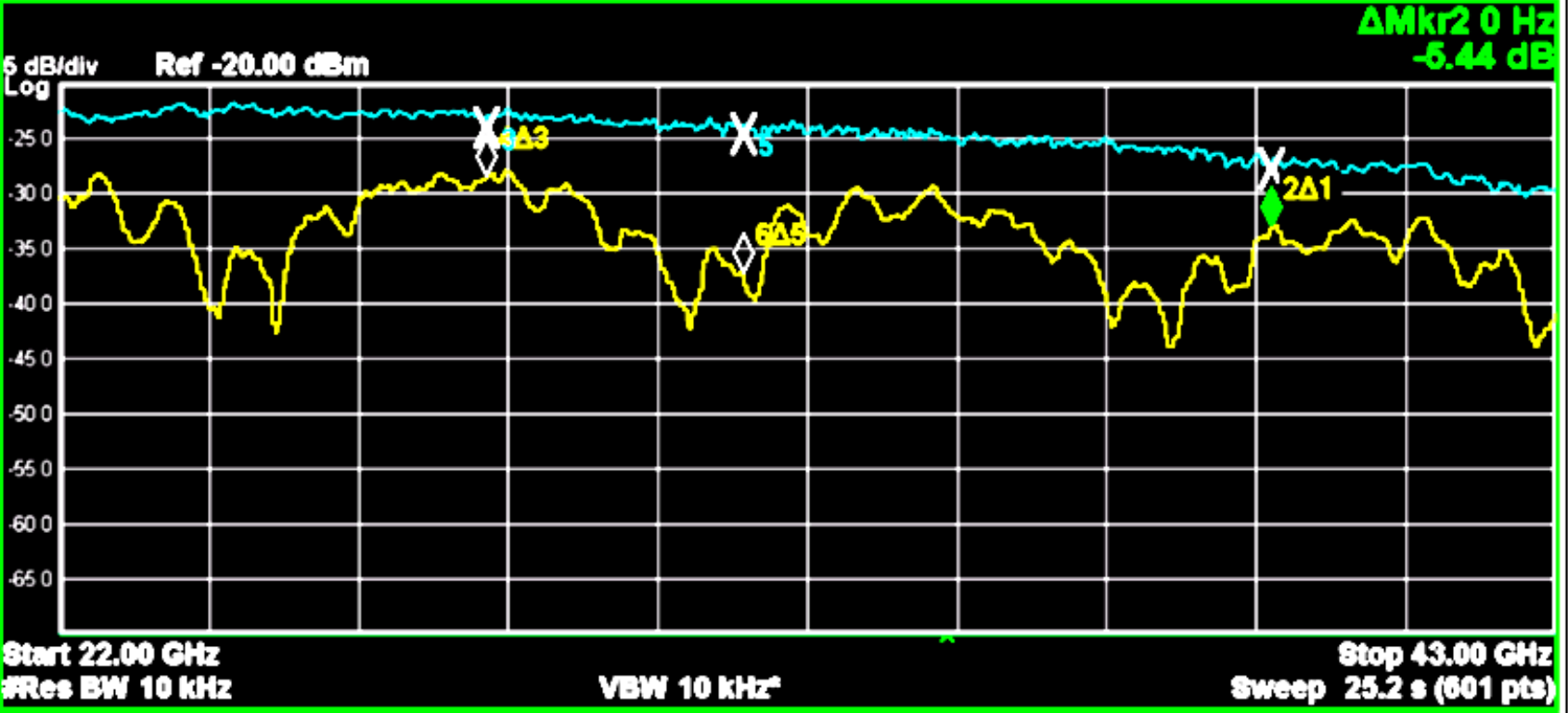}
\\ {\hspace{0.35in}} (c) & {\hspace{0.12in}} (d)
\end{tabular}
\caption{\label{fig_mat}
Material measurements illustrating reflection response over sheetrock using (a) parallel
and (b) perpendicular polarizations. 
Response with (c) a structured partition wall, and (d) an external wall in the Qualcomm building
across a range of mmW frequencies.}
\end{center}
\vspace{-5mm}
\end{figure*}

Reflection tests were conducted with different materials across a large range of
incidence angles and for both parallel and perpendicular polarizations. For the
sake of illustration, Figs.~\ref{fig_mat}(a)-(b) illustrate the 
reflection response with a $5/8$ inch sheetrock material over the $22$-$43$ GHz range at parallel and
perpendicular polarizations with an incidence angle of $18.5^{\sf o}$. The main
observation here is that periodic notches that are several GHz wide and often with
more than $30$ and $35$ dB in loss, respectively, are seen. These losses are attributed
to changing material properties with frequency due to which signals constructively/destructively
interfere from different surfaces that make the material. While a similar trend is observed
across these experiments for both polarizations and different choices of incidence angles,
the precise response at a frequency and the depth of the notches depend on the material, incidence
angle and polarization. 

Fig.~\ref{fig_mat}(c) shows the more complicated (but realistic) response of a structured
partition wall with multiple layers of materials (two sheetrock plates separated by a $4$
inch air gap) at $18.5^{\sf o}$ incidence angle and perpendicular polarization. The
superposition of the response from the individual layers leads to complicated/periodic
patterns across the frequency range (yellow curve), whereas the response of the single
sheetrock alone is presented in the green curve. Fig.~\ref{fig_mat}(d) illustrates
the reflection response with a typical external wall material in the Qualcomm building,
which is similar in behavior as Fig.~\ref{fig_mat}(c).

\begin{figure*}[htb!]
\begin{center}
\begin{tabular}{cc}
\includegraphics[height=2.1in,width=2.4in] {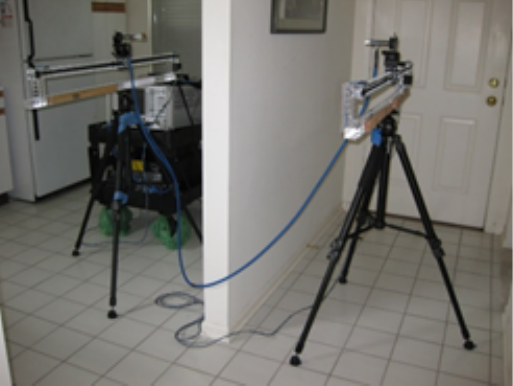}
&
\includegraphics[height=2.3in,width=3.1in] {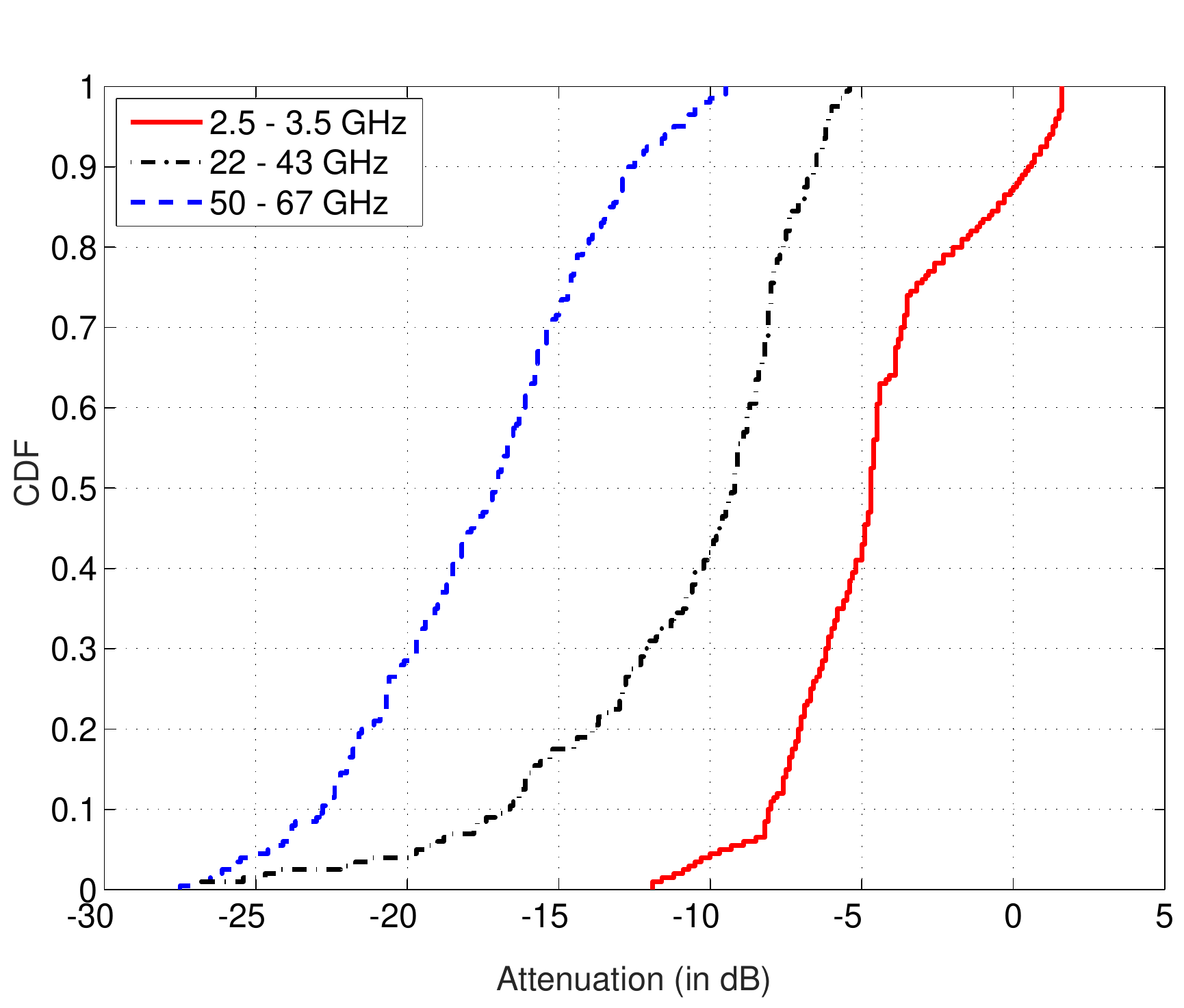}
\\
(a) & (b)
\end{tabular}
\caption{\label{fig_penetration}
(a) Penetration loss measurement setup.
(b) CDF of penetration loss with exterior walls in residential buildings.
}
\end{center}
\end{figure*}

\subsection{Penetration Loss}
We now consider studies with common residential wall materials to understand the scope of
penetration loss for outdoor-to-indoor coverage and residential deployments. The measurement
setup consists of two horn antennas (both with $10$ dBi gain) pointing at each other through
the wall at a normal angle of incidence as in Fig.~\ref{fig_penetration}(a). The distance
from the wall is about $0.5$ m for either antenna. For the measurements,
the antennas are moved along and parallel to the wall with the receive antenna tracking the
position of the transmit antenna in steps of $\sim 2.75$ mm over a $1$ m distance. The tests
use a broadband sweep over $22$-$43$ GHz and $50$-$67$ GHz using a vector network analyzer and
horn antennas. A broadband sweep is necessary to mitigate multi-surface reflections in the drywall
and the complex wall structure. For reference, omni antenna measurements over the $2.5$-$3.5$ GHz
range are also obtained. 

From our studies, we first note that sheathing material made of strand boards
(wood chips) involve the heavy use of glue, which has more attenuation
at higher frequencies. This is reflected in Fig.~\ref{fig_penetration}(b) which
illustrates the CDF of penetration loss with a strand board construction. In
particular, median values of $4.7$, $9.2$ and $17.1$ dB are observed in the $2.5$-$3.5$,
$22$-$43$ and $50$-$67$ GHz regimes, respectively. For walls made of plywood material,
smaller losses that are comparable over a wide frequency range are observed. In particular,
median values of $2.2$ and $3.0$ dB are seen in the $22$-$43$ and $50$-$67$ GHz regimes,
respectively.
Since strand board is typically lower cost than plywood, it is likely that newer/urban
constructions as well as exterior residential walls are more likely to use strand board
material than plywood material~\cite[p.\ 5]{apawood}. Interior residential walls are
more likely to be made of plywood material. Thus, it appears that outdoor-to-indoor coverage is
more likely in older/residential settings than in urban settings.

\begin{figure*}[htb!]
\begin{center}
\includegraphics[height=3.8in,width=6.15in] {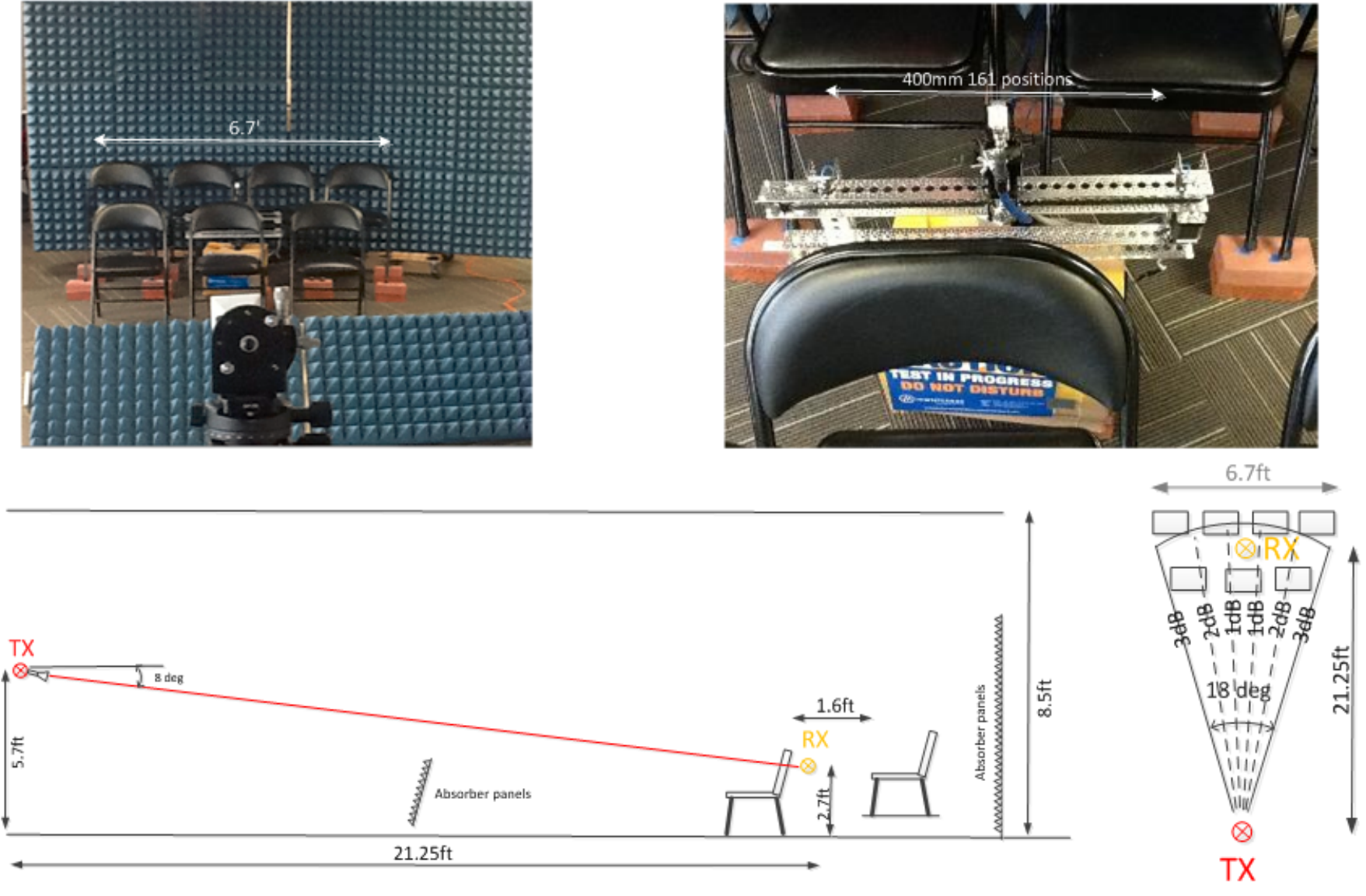}
\caption{
Layout of the stadium emulation experiments along with the linear
measurement track, side view and top view of the setup.
}
\label{fig_stadium}
\end{center}
\vspace{-5mm}
\end{figure*}

\subsection{Stadium Emulation}
Another important aspect that needs understanding is the reflection/penetration of
mmW signals from/through the human body. To understand this aspect, an experiment 
with two rows of improvised seats emulating a stadium setting is performed. The
experimental setup is as shown in Fig.~\ref{fig_stadium} with four people seated
in the back row and three people in the front row. The back row is elevated over
the front by four inches with the seats arranged in a staggered fashion. Chairs
are made of metal covered with vinyl cushion. Absorbing panels are placed behind
the back row and ground bounce is mitigated by absorbing panels on the floor.

\begin{figure*}[htb!]
\begin{center}
\includegraphics[height=2.3in,width=3.1in] {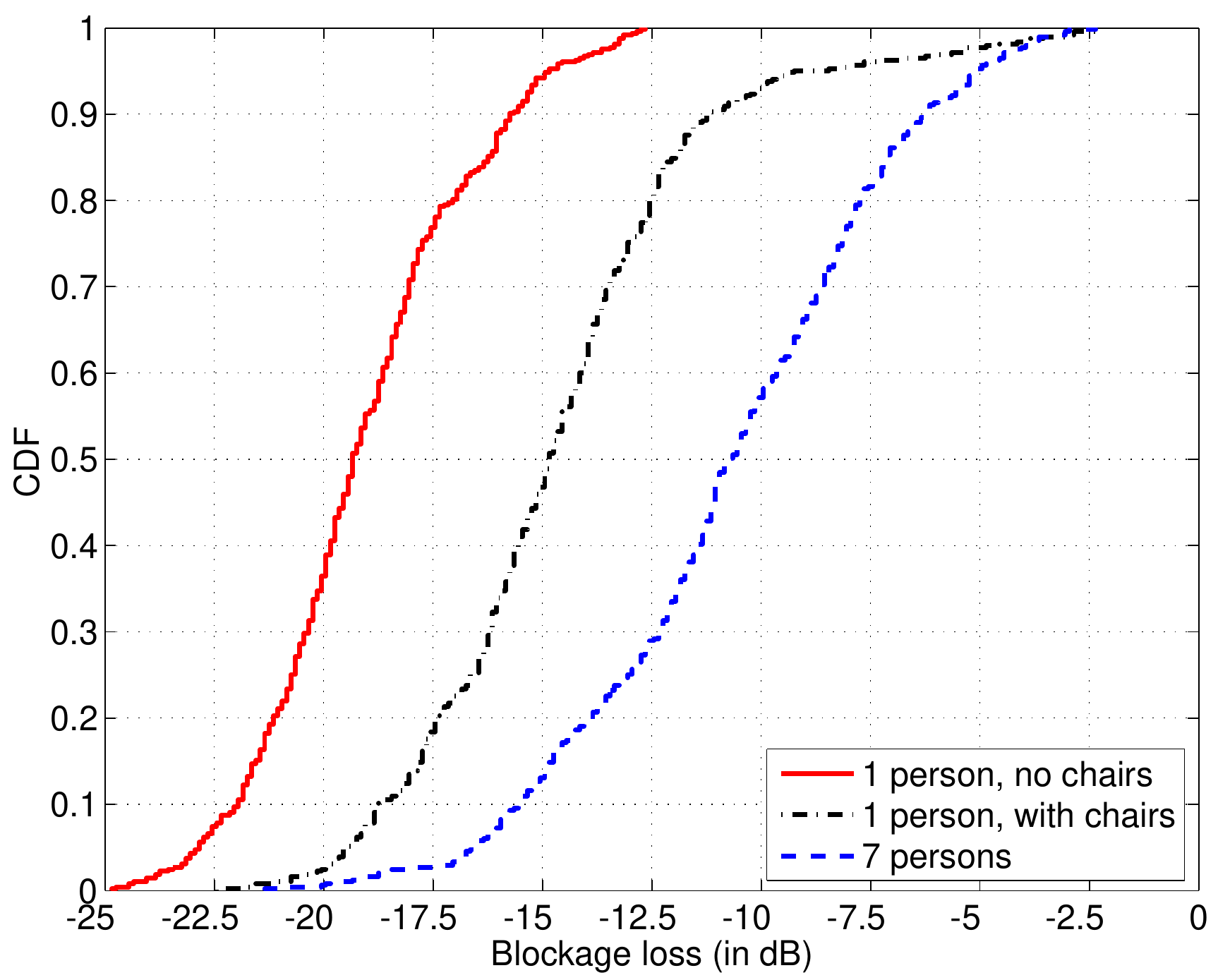}
\caption{\label{fig_blockage}
CDF of blockage loss in a stadium emulation.
}
\end{center}
\end{figure*}

A horn antenna with $20$ dBi gain and a beamwidth of $15^{\sf o}$ is placed at a
distance of $6.5$ m, a height of $1.7$ m and at a downtilt angle of $98^{\sf o}$.
An omni-directional antenna is placed behind the middle seat of the front row
at $\approx {\hspace{-0.04in}} 4$ inches behind the first row and multiple
measurements are made at $29$ GHz. The antenna is moved on a linear track
every $\frac{1}{4}$ wavelength over $161$ positions (approx.\ $40$ cm) and
measurements are obtained. Based on these measurements, it is observed that in
general, human body scatters
energy on to nearby geographic locations thereby aiding in stadium deployments by
providing secondary bounces (alternate paths) for signaling when the LOS path in
the boresight direction is blocked. To understand this observation, four controlled
experiments are performed: i) no persons and no chairs (for baseline/reference),
ii) one person with no surrounding chairs, iii) one person with all the surrounding chairs,
and iv) seven persons in their respective chairs. Blockage loss is computed for the
three latter scenarios relative to the baseline scenario of no persons and no chairs.
Fig.~\ref{fig_blockage} illustrates the CDF of
blockage loss 
in these three scenarios. 
In particular, the effect of adjacent chairs reduces the blockage loss
significantly from a median of $19.2$ dB to $14.9$ dB, and the presence of nearby
humans improves
the median further to $10.8$ dB. Thus, this study illustrates that these additional
reflections and energy accumulation need to be modeled in stadium use-cases.


\section{Implications on PHY Layer Design}
\label{sec6}
Based on the measurements described in Sections.~\ref{sec3}-\ref{sec4}
and comparisons of different figures-of-merit at identical
transmit-receive location pairs across different frequencies, we make the
following observations on the implications of these measurements on PHY layer/system design.
\begin{enumerate}
\item
Indoor office, mall and outdoor deployments
suggest the viability of good (but frequency-dependent) link margins with LOS links
in a moderate number of cases. In scenarios with a far-field obstructed LOS path
leading to a NLOS link, frequency-dependent shadow fading and path losses are
observed. Losses at mmW frequencies (with respect to a reference distance of $d_0 = 1$ m)
are typically higher than with sub-$6$ GHz systems in both indoor and outdoor settings.
Nevertheless, the differential impact of mmW systems in terms of link margins
relative to sub-$6$ GHz systems is {\em not} dramatically worse.

\item
While we do not report these studies here, our initial work~\cite{qualcomm3}
as well as other similar
works~\cite{802d11_maltsev,peter_et_al,metis2020,maccartney_vtc2016,maccartney_2017}
in the literature (also, see the 3GPP Rel.\ 14 channel modeling
document~\cite[pp.\ 49-52]{3gpp_CM_rel14}) show that significant additional
impairments due to human/hand blockages are observed at mmW frequencies.

\item
Other works such as~\cite{wang2,jones_report} as well as our internal
measurements point
at significant losses due to foliage. In the outdoor-to-indoor coverage scenario,
reflection response and penetration loss that are a function of material property, frequency, polarization
and incident angle is observed. Further, deep notches that are several GHz wide
are also seen. This motivates the need for system designs that support {\em both}
frequency and spatial diversity.

\item
The path, penetration and blockage losses at mmW frequencies along with typical
equivalent isotropically radiated power (EIRP) constraints suggest that low
pre-beamforming ${\sf SNR}$s are the norm in mmW systems. Thus, a viable system
design has to overcome these huge losses. These losses can be bridged with
beamforming array gains from the packing of a large number of antennas within the
same array
aperture~\cite{hur,roh,brady,sun,oelayach,raghavan_jstsp,rangan}. 
Given the energy and complexity tradeoffs associated with large arrays, typical
antenna geometries at the base-station are $64 \times 4$, $32 \times 8$, etc., with
$2$-$16$ layer transmission.

\item
In particular, subarray diversity is critical to overcome near-field
obstructions such as those due to different parts of the human body that
can significantly impair the received signal quality. This is also important
to ensure coverage at the UE side over the entire angular space. 
Due to the smaller $\lambda$ at $60$ GHz
(relative to $28$ GHz), more subarrays can be packed in the same area and
such capabilities should be leveraged for better performance. While a large number
of subarrays can be envisioned in a UE design, cost and complexity considerations
suggest the use of $2$-$4$ layers with each layer independently controlling a
subarray of $2$-$8$ antennas. 

\item

Clustering of multipath~\cite{fleury,czink} is an important macroscopic property
that needs to be understood since the dominant clusters/angles capture the modes of
propagation and are hence useful/relevant for multi-layer beamforming in transmission.
This is a topic requiring careful attention and future work will address this aspect
in detail. We point the readers to initial work in~\cite{qualcomm2} which show that
on average, 
$4$-$5$ clusters appear to be within a power differential of $5$ dB of each other
motivating both directional transmissions and suggesting a high level of diversity.
The viability of multiple modes suggests the use of both single-user MIMO
strategies for increasing the peak rate as well as multi-user MIMO strategies
for increasing the sum-rate~\cite{sun,vasanth_gcom16}.

\item
Practical beamforming algorithms should simultaneously optimize multiple criteria
such as: i) good beamforming gain, ii) less unintended interference, iii) a link
margin-dependent hierarchical solution for beamformer learning allowing a smooth
tradeoff between beamforming gain and number of
training samples, 
iv) robustness to channel dynamics, v) ability to work with different beamforming
architectures, vi) a simpler network architecture that allows for a broadcast
solution in initial UE discovery, etc.
Coupled with higher antenna dimensionality in the mmW regime, the {\em sparse}
and {\em directional} channel structure motivates the use of a certain subset
of {\em directional} beamforming strategies~\cite{oelayach,raghavan_jstsp,brady,vasanth_gcom15,vasanth_gcom16,vasanth_jsac2017}. %
Such strategies allow a tradeoff between peak beamforming gain and initial UE
discovery latency~\cite{raghavan_jstsp,barati,ghadikolaei}.

\item
Given the small wavelengths, robust beam tracking is necessary to maintain the
link gains even though the path parameters (gains and phases) change at much
smaller time-scales. Further, given the significantly reduced interference due
to a directional beam in unintended directions, the channel structure also
motivates the use of simpler directional schedulers for user scheduling allowing
dense spatial reuse and thus a higher network capacity~\cite{vasanth_gcom16,vasanth_jsac2017}.

\item
Frequency-dependent delay spreads are observed in NLOS settings with both
omni-directional scans and directional
beamforming. While delay spreads are 
small in most scenarios (for example, they are on the order of $30$-$50$ ns in
indoor office, $50$-$90$ ns in shopping mall and $150$-$300$ ns in outdoor
street canyon-type settings), there are also scenarios where a significantly large delay
spread is seen (for example, even up to $800$ ns in certain outdoor open square
settings with omni antennas).
These extremes in outdoor settings can be attributed to 
radar cross-section effect. 
Supporting these extremes without
incurring a high fixed system overhead (in terms of the cyclic prefix length for a
multi-carrier design) is important.

\item
In addition to the likelihood of multiple viable paths to a certain base-station,
there are also viable paths to multiple base-stations~\cite{jung,vasanth_comm_mag_16}.
While the observation in~\cite{jung} is based on ray-tracing,~\cite{vasanth_comm_mag_16}
is based on field measurements in both indoor and outdoor mobility studies.
These observations suggest the criticality of a dense deployment of base-stations for robust mmW
operation~\cite{qualcomm} and inter-base-station handover to leverage these paths.
Integrated access and backhaul operation is highly desirable for small cell
deployment~\cite{qualcomm}.

\item
Measurements emulating a stadium environment indicate that the presence of nearby
humans can 
improve the received signal quality. These observations
suggest the importance of not only modeling human blockage, but also reflection
and energy accumulation from humans in such scenarios. 
These aspects require further investigation to address specular reflection and absorption.



\end{enumerate}

\section{Concluding Remarks}
\label{sec7}

The main focus of this paper has been on a comparative study of propagation
at $2.9$, $29$ and $61$ GHz across a large number of transmit-receive location
pairs over different propagation environments (indoor office, shopping mall,
outdoor settings, etc.). Our studies show that path loss at millimeter wave
frequencies varies only less substantially relative to path loss at sub-$6$ GHz
bands. On the other hand, delay spreads across LOS and NLOS links could vary
substantially across frequencies. Furthermore, key differences are observed in
terms of penetration of electromagnetic radiation across different materials
commonly found in an indoor/residential setting. Deep notches in frequency
response due to material properties suggest the use of frequency and spatial
diversity schemes for communications. Human body scatters millimeter wave
radiation to nearby geographic locations enhancing scenarios such as stadium
coverage. 

Observations from the studies described in this paper have already had a
fundamental impact on channel models for $> {\hspace{-0.03in}} 6$ GHz systems.
Studies on path loss, delay spread, material measurements, etc.\ have played a
significant part in the multi-institutional channel modeling effort for
$> {\hspace{-0.03in}} 6$ GHz systems~\cite{5G_whitepaper}. 
The channel measurements are
also used to obtain key system design guidelines on UE structure/geometry/design,
beamforming, waveforms, etc.\ and these guidelines are important given the
{\em accelerated schedule} of the
Fifth Generation New Radio (5G NR) process at 3GPP and beyond. Results from the
cited measurement campaigns highlight differences as strongly as
consensus (as noted from~\cite{5G_whitepaper}),
illustrating the need for further extensive measurements in diverse settings. Of
particular interest are more measurements on material propagation, blockage modeling,
stadium use-case modeling, as well as simpler stochastic models that capture these
measurements from a system level design perspective.

\bibliographystyle{IEEEtran}
\bibliography{newrefsx}

\end{document}